\documentclass[12pt,fleqn]{article}
\usepackage{epsf}
\renewcommand{\baselinestretch}{1.2}
\makeatletter
\def\section{\@startsection {section}{1}{\z@}{3.5ex plus 1ex minus
    .2ex}{2.3ex plus .2ex}{\sc }}
\def\subsection{\@startsection{subsection}{2}{\z@}{3.25ex plus 1ex
minus
   .2ex}{1.5ex plus .2ex}{\small \sc }}
\def\appendix{\par\clearpage
  \setcounter{section}{0}
  \setcounter{subsection}{0}
  \@addtoreset{equation}{section}
  \def\@sectname{Appendix~}
  \def\theequation{\thesection.\arabic{equation}}
  \def\thesection{\Alph{section}}}
\makeatother
\makeatletter \@addtoreset{equation}{section} \makeatother
\renewcommand{\theequation}{\thesection.\arabic{equation}}

\def\ap#1#2#3{     {\it Ann. Phys. (NY) }{\bf #1} (19#2) #3}

\def\npb#1#2#3{    {\it Nucl. Phys. }{\bf B #1} (19#2) #3}
\def\plb#1#2#3{    {\it Phys. Lett. }{\bf B #1} (19#2) #3}
\def\prd#1#2#3{    {\it Phys. Rev. }{\bf D #1} (19#2) #3}

\def\prl#1#2#3{    {\it Phys. Rev. Lett. }{\bf #1} (19#2) #3}

\def\zpc#1#2#3{    {\it Z. Phys. }{\bf C #1} (19#2) #3}

\def\nc#1#2#3{     {\it Nuovo Cim. }{\bf #1} (19#2) #3}

\def\ijmpa#1#2#3{  {\it Int. J. Mod. Phys. }{\bf A #1} (19#2) #3}

\def\eq#1{{eq.~(\ref{#1})}}
\def\eqs#1#2{{eqs.~(\ref{#1})--(\ref{#2})}}

\let\vev\VEV
\def\abs#1{\left| #1\right|}
\def\mod#1{\abs{#1}}
\def\Im{\mathop{\mbox{Im}}}
\def\Re{\mathop{\mbox{Re}}}

\def\etal{{\it et al.}}

\newcommand{\bea}{\begin{eqnarray}}
\newcommand{\beq}{\begin{equation}}
\newcommand{\eea}{\end{eqnarray}}
\newcommand{\eeq}{\end{equation}}
\newcommand{\nnu}{\nonumber}
\newcommand{\spav}[1]{\parbox{1mm}{\vspace*{#1}}}

\begin{document}
\begin{titlepage}
\vspace*{-1cm}
\begin{flushright}
{\tt SISSA 103/95/EP}
\end{flushright}
\spav{0.0cm}
\begin{center}
{\Large\bf A New Estimate of $\varepsilon '/\varepsilon$}\\
\spav{0.7cm}\\
{\large S. Bertolini$^{\dag}$, J.O. Eeg$^{\ddag}$ and
M. Fabbrichesi$^{\dag}$} 
\spav{1.0cm}\\
{\em  $^{\dag}$ INFN, Sezione di Trieste, and}\\
{\em Scuola Internazionale Superiore di Studi Avanzati}\\
{\em via Beirut 4, I-34013 Trieste, Italy.}\\

{\em $^{\ddag}$ Department of Physics, University of Oslo}\\
{\em N-0316 Oslo, Norway.}\\
\spav{1.0cm}\\
{\sc Abstract}
\end{center}
{\footnotesize
We discuss direct $CP$ violation in the standard model by
giving  a new  estimate 
of $\varepsilon '/\varepsilon$ in kaon decays. Our analysis
is based on  the evaluation of the 
hadronic matrix elements of the  \mbox{$\Delta S =1$} 
effective quark lagrangian
by means of the chiral quark model, with the inclusion of
meson one-loop renormalization and NLO Wilson coefficients. 
Our estimate is fully consistent with the $\Delta I =1/2$
selection rule in $K\to \pi\pi$ decays which is well reproduced within the
same framework.
By varying all parameters in the allowed ranges and, in
particular, 
taking the quark condensate---which is the major source of
uncertainty---between $(-200\ {\rm MeV})^3$ and $(-280\ 
{\rm MeV})^3$
we find
$$ -5.0 \times 10^{-3}\ < \varepsilon '/\varepsilon 
   < \ 1.4 \times 10^{-3} \, .$$
Assuming for the quark condensate the improved
PCAC result
\mbox{$\vev{\bar qq} = -(221\: \pm 17\ {\rm MeV})^3$} 
 and fixing $\Lambda_{\rm QCD}^{(4)}$ to its central value,  we find the more
restrictive prediction
$$\varepsilon '/\varepsilon = ( 4 \pm 5 ) \,\times \,10^{-4}\ , $$
where the central value is defined as the average over the allowed
values of Im $\lambda_t$ in the first and second quadrants.
In these estimates
the relevant mixing parameter Im $\lambda_t$ is self-consistently
obtained from
$\varepsilon$ and we take
$m_t^{\rm pole} = 180 \pm 12$ GeV.
Our result is, to a very good approximation,
 renormalization-scale and $\gamma_5$-scheme independent. 
 }
\vfill
\spav{.5cm}\\
{\tt SISSA 103/95/EP}\\
{\tt  November 1995 }

\end{titlepage}

\newpage
\setcounter{footnote}{0}
\setcounter{page}{1}

\section{Introduction}

The real part of $\varepsilon'/\varepsilon$ measures 
direct $CP$ violation in the decays of a neutral kaon in two 
 pions. It is a fundamental quantity which has justly attracted a great
 deal of theoretical as well as experimental work. Its
 determination would answer the question of whether
 $CP$ violation is 
 present only 
 in the mass matrix of neutral kaons (the superweak scenario) or is instead
 at work also directly in the decays. 
 
 On the experimental
 front, the present results of CERN (NA31)~\cite{CERN}
 \beq
 \mbox{Re} \: \left( \varepsilon'/\varepsilon \right) = 
 (23 \pm 7) \times 10^{-4}
 \eeq
 and Fermilab (E731)~\cite{SLAC}
 \beq
 \mbox{Re} \: \left( \varepsilon'/\varepsilon \right) = 
 (7.4 \pm 6.0) \times 10^{-4}
 \eeq
are tantalizing insofar as the superweak scenario cannot be excluded and
the disagreement between the two outcomes still leaves a large 
uncertainty. The next generation of experiments---presently under way at
CERN, Fermilab and DA$\Phi$NE---will
improve the sensitiveness to $1 \times 10^{-4}$ and hopefully reach
 a definite result.

On the theoretical side, much has been accomplished, although the 
intrinsic difficulty
of a problem that encompasses scales as different as $m_t$ and $m_\pi$
weights against any decisive progress in the field.

A fundamental step was recently covered by the 
Munich~\cite{monaco}
and Rome~\cite{roma} groups who computed the anomalous dimension matrix
of the ten
relevant operators to the next-to-leading order (NLO) in two 
$\gamma_5$-schemes of dimensional regularization: 't Hooft-Veltman (HV) and
Naive Dimensional Regularization (NDR). This computation has brought the 
short-distance part of the effective lagrangian under control.

The residual (and, unfortunately, largest) uncertainty is
 due to the long-distance part of the lagrangian, the computation of which
 implies evaluating the
hadronic matrix elements of the quark operators. It is here that the
non-perturbative regime of QCD is necessarily present and our understanding
is accordingly blurred.
 
At present,
there exist two complete estimates of such hadronic matrix elements 
performed by the aforementioned
groups, and recently updated in ref.
\cite{martinelli} for the lattice (at least for some of
the operators), where the value 
\beq
 \mbox{Re} \: \left( \varepsilon'/\varepsilon \right) = 
 (3.1\pm 2.5 \pm 0.3 ) \times 10^{-4} \label{estI}
 \eeq
is found, and ref.~\cite{buras} 
for the $1/N_c$ approach (for all ten operators) improved by
fitting the $\Delta I = 1/2$ rule, where 
$\varepsilon'/\varepsilon$ is estimated to be within the range
\beq
 - 2.5 \times 10^{-4} \leq \mbox{Re} \: \left( \varepsilon'/\varepsilon \right)
 \leq 13.7 \times 10^{-4} \, .\label{estII}
 \eeq 

The smaller error in \eq{estI} originates in the Gaussian treatment of
the uncertainty in the input parameters 
with respect to the flat 1$\sigma$ error
included in \eq{estII}.

 Both groups seem to agree on the difficulty of accommodating
 within the standard model a value substantially
 larger than $ 1 \times 10^{-3}$. This unexpectedly small value is the result of
the cancellation between gluon and electroweak penguin 
operators~\cite{FR}. If that is actually the case, it is somewhat
disappointing that 
  the presence of direct $CP$ violation in the standard model
 turns out to be hidden by an accidental cancellation that
 effectively mimics the superweak scenario.

It seemed to us that a third, independent estimate of
$\varepsilon'/\varepsilon$ was  
desirable and we have taken the point of view
that a reliable evaluation of the hadronic matrix elements
 should first provide a consistent 
picture of kaon physics, starting
from the $CP$-conserving amplitudes and, in particular, by reproducing
the $\Delta I = 1/2$ selection rule, which governs most of these
amplitudes as well as the quantity 
$\varepsilon'/\varepsilon$ itself. 
We also felt that the same evaluation should pay particular
attention to the problem of achieving a satisfactory 
$\gamma_5$-scheme and scale independence in the matching between the
matrix elements and the Wilson coefficients, the absence of which 
would undermine any estimate.

In a preliminary work~\cite{BEF}, we studied 
$\varepsilon'/\varepsilon$
within the chiral quark model ($\chi$QM)~\cite{QM} in a toy model
that included the leading effect of the two most important operators,
and verified that
the $\gamma_5$-scheme independence could be  achieved.

In a recent paper~\cite{ABEFL}, hereafter referred as I, we have completed the
study of the hadronic matrix elements of all the ten
operator of the $\Delta S =1$ effective quark
langrangian by means of the $\chi$QM and verified in~\cite{ABFL}, hereafter
referred as II, that the
inclusion of non-perturbative $O(\alpha_s N_c)$ corrections and
one-loop meson renormalization provided an
improved scale independence
and, more importantly, a good fit of the $\Delta I =1/2$ selection rule. 

These results put us
in the position to provide a new estimate of $\varepsilon'/\varepsilon$
that is
independent of the existing ones and that contains new features that, 
in our judgment, makes it more reliable.

We summarize here such features.
Our estimate takes advantage, as the existing ones, of
\begin{itemize}
\item
NLO results for the Wilson coefficients;
\item 
up-to-date analysis of the constraints on the mixing coefficient
Im $\lambda_t$.
\end{itemize}
Among the new elements introduced, the most relevant are 
\begin{itemize}
\item
A consistent evaluation of all hadronic matrix elements in the $\chi$QM
(including non-perturbative gluon condensate effects) in two schemes
of dimensional regularization;
\item
Inclusion in the $\Delta S=1$ chiral lagrangian of the
complete bosonization $O(p^2)$ of the electroweak operators $Q_7$ and
$Q_8$. Some relevant $O(p^2)$ terms have been neglected
in all previous estimates;
\item
Inclusion of the meson-loop renormalization and scale dependence
of the matrix elements;
\item
Consistency with the $\Delta I = 1/2$ selection rule in kaon decays;
\item
Matching-scale and $\gamma_5$-scheme dependence of the results below the 
20\% level. 
\end{itemize}

Even though our framework enjoys a high degree of reliability, 
any estimate of $\varepsilon'/\varepsilon$ necessarily
suffers of a systematic uncertainty that cannot be
easily  reduced further. We
find that it is mainly parameterized in terms of the value of 
the quark condensate,
the input parameter that dominates penguin-diagram physics. 
For this reason, we  discuss first a inclusive estimate based on a conservative
 range of $\vev{\bar{q}q}$, as well as the variations of all the
other inputs: $m_t$,  $\Im \lambda_t$ (which depends, beside $m_t$ and
$m_c$, on
$\hat B_K$ and other mixing angles) and 
$\Lambda_{\rm QCD}$.
 Such a procedure provides us with the range of
values for $\varepsilon'/\varepsilon$
that we consider to be the unbiased theoretical prediction of the standard model.
Unfortunately, this range turns out to be  rather large, spanning, as it can
be seen in the abstract, from $-5 \times 10^{-3}$ to $1.4 \times 10^{-3}$. On the
other hand, it is as small as we can get without making some further
assumptions on the input parameters---assumptions that all the
other available estimates must make as well.

In order to provide such a more restrictive estimate,  we have 
   chosen the improved  PCAC prediction for
 the quark condensate and fixed $\Lambda_{\rm QCD}^{(4)}$ to its
central value.  This
reasonable, but nevertheless arbitrary choice
 allows us to give the
second, and more predictive estimate reported in the abstract. It is the latter
 that
should be compared with the current estimates, while, at the same time, bearing
in mind also the former unrestricted range as a realistic measure of our
ignorance.

Such uncertainty notwithstanding, we agree in the end with the main point of
ref.~\cite{martinelli}, namely that it is difficult to accommodate 
within the standard
model a value of $\varepsilon'/\varepsilon$ larger than $1 \times 10^{-3}$.
In fact, if our analysis points toward a definite prediction, it points to even
smaller values, if not negative ones. This can be understood not so much as
a peculiar feature of the $\chi$QM prediction as the neglect in other estimates
of a class
of contributions in the vacuum saturation approximation (VSA)
 of the matrix elements of the electroweak operators. This problem
 is discussed in detail
in I. These new contributions
 are responsible for the  onset of the superweak regime for values of $m_t$
less than 200 GeV.
In our computation, it is  the meson renormalization that in the end brings back  
$\varepsilon'/\varepsilon$ around zero or positive values. 

The outline of the paper is the following. In section 2 we write the effective
quark lagrangian, discuss the short-distance input parameters and give the
Wilson coefficients. Section 3 contains a brief discussion of the $\chi$QM 
evaluation of the hadronic matrix elements and their corresponding meson-loop
renormalization. In section 4  we discuss the values of the
input parameters and in section 5 the effective factors $B_i$'s that give
the comparison between the VSA and the
$\chi$QM evaluation of the hadronic matrix elements. We begin in
section 6 our discussion of
 $\varepsilon'/\varepsilon$  by first studying the $\gamma_5$-scheme
independence and then the contribution of each operator taken by itself. In
section 7, we give our estimate as a function of the most important input
parameters in a series of figures and one table. The numerical value of
all input parameters  are collected in a table in the appendix. 

 \section{The Quark Effective Lagrangian and the NLO Wilson Coefficients}

 The quark effective lagrangian at a scale $\mu < m_c$ can be written
as~\cite{GW}
 \bea
{\cal L}_{\Delta S = 1} &=& 
-\frac{G_F}{\sqrt{2}} V_{ud}\,V^*_{us} \sum_i \Bigl[
z_i(\mu) + \tau y_i(\mu) \Bigr] Q_i (\mu) \nnu \\ 
&\equiv& -\frac{G_F}{\sqrt{2}} \sum_i C_i (\mu) Q_i (\mu)  
\, . 
\label{ham}
\eea

The $Q_i$ are four-quark operators obtained by integrating out in the standard
model the vector bosons and the heavy quarks $t,\,b$ and $c$. A convenient
and by now standard
basis includes the following ten quark operators:
 \beq
\begin{array}{rcl}
Q_{1} & = & \left( \overline{s}_{\alpha} u_{\beta}  \right)_{\rm V-A}
            \left( \overline{u}_{\beta}  d_{\alpha} \right)_{\rm V-A}
\, , \\[1ex]
Q_{2} & = & \left( \overline{s} u \right)_{\rm V-A}
            \left( \overline{u} d \right)_{\rm V-A}
\, , \\[1ex]
Q_{3,5} & = & \left( \overline{s} d \right)_{\rm V-A}
   \sum_{q} \left( \overline{q} q \right)_{\rm V\mp A}
\, , \\[1ex]
Q_{4,6} & = & \left( \overline{s}_{\alpha} d_{\beta}  \right)_{\rm V-A}
   \sum_{q} ( \overline{q}_{\beta}  q_{\alpha} )_{\rm V\mp A}
\, , \\[1ex]
Q_{7,9} & = & \frac{3}{2} \left( \overline{s} d \right)_{\rm V-A}
         \sum_{q} \hat{e}_q \left( \overline{q} q \right)_{\rm V\pm A}
\, , \\[1ex]
Q_{8,10} & = & \frac{3}{2} \left( \overline{s}_{\alpha} 
                                                 d_{\beta} \right)_{\rm V-A}
     \sum_{q} \hat{e}_q ( \overline{q}_{\beta}  q_{\alpha})_{\rm V\pm A}
\, , 
\end{array}  
\label{Q1-10} 
\eeq
where $\alpha$, $\beta$ denote color indices ($\alpha,\beta
=1,\ldots,N_c$) and $\hat{e}_q$  are quark charges. Color
indices for the color singlet operators are omitted. 
The labels \mbox{$(V\pm A)$} refer to 
\mbox{$\gamma_{\mu} (1 \pm \gamma_5)$}.
We recall that
$Q_{1,2}$ stand for the $W$-induced current--current
operators, $Q_{3-6}$ for the
QCD penguin operators and $Q_{7-10}$ for the electroweak penguin (and box)
ones. 

The functions $z_i(\mu)$ and $y_i(\mu)$ are the
 Wilson coefficients and $V_{ij}$ the
Koba\-ya\-shi-Mas\-kawa (KM) matrix elements; $\tau = - V_{td}
V_{ts}^{*}/V_{ud} 
V_{us}^{*}$. Following the usual parametrization of the KM matrix, in order to
determine $\varepsilon'/\varepsilon$, we only
need the $y_i(\mu)$, which control the $CP$-violating part of the amplitudes.

The size of the Wilson coefficients at the hadronic scale ($\sim 1$ GeV) 
depends on $\alpha_s$ and the threshold masses $m_W$, $m_b$ and $m_c$. 
In addition, the penguin coefficients $y_i(\mu)$ depend on the top mass
via the initial matching conditions.

The  
recent determination of the strong coupling at LEP and SLC gives~\cite{alfa}
\beq
\alpha_s (m_Z) = 0.119 \pm 0.006 \, ,
\eeq
which corresponds to
\beq
\Lambda^{(4)}_{QCD} = 350 \pm 100 \: \mbox{MeV} \, .
\label{lambdone}
\eeq
We will use the range in \eq{lambdone} for our numerical estimate of 
$\varepsilon'/\varepsilon$.

For $m_t$ we take the value~\cite{PDG}
\beq
m_t^{\rm pole} = 180 \pm 12 \:\:\mbox{GeV} 
\label{mtpole}
\eeq
The relation between the pole mass $M$ and the $\overline{\rm MS}$ 
running mass $\overline{m}(\mu)$ 
is given at one loop in QCD by~\cite{defpole}: 
\beq
\overline{m}(M)\ =\ 
   {M(q^2=M^2)}\ {\left(1-\frac{4}{3}\frac{\alpha_s(M)}{\pi}\right) }
\label{polemass} \ \ ,
\eeq
For the running top quark mass, in the range of $\alpha_s$
considered, we obtain
\beq
 \overline{m}_t(m_t^{\rm pole})  \simeq 172 \pm 12 \:\:\mbox{GeV}
\label{mt-mt}
\eeq
which, using the one-loop running, corresponds to 
\beq
 \overline{m}_t(m_W)  \simeq 183 \pm 14 \:\:\mbox{GeV} \, ,
\label{mt-mw}
\eeq
which is the value to be used as input at the $m_W$ scale for the 
NLO evolution of the Wilson coefficients. In \eq{mt-mw} we have averaged over
the range of $\Lambda^{(4)}_{QCD}$ given in \eq{lambdone}. 
We have explicitly checked that taking $\mu=m_t^{\rm pole}$ as the
initial matching scale in place of $m_W$
and using correspondingly $\overline{m}_t(m_t^{\rm pole})$ the  
electroweak Wilson coefficients at $\mu= 1$ GeV remain stable up to
the percent level, 
while the variation of the relevant gluon penguin coefficients 
stays below 15\%. The stability worsen by keeping the top mass
fixed, while varying the matching scale.

For $m_b$  we take the value
\beq
m_b^{\rm pole} = 4.8 \:\:\mbox{GeV}
\label{mbpole}
\eeq
which falls in the range $4.5 - 4.9$ given in \cite{PDG},
and for $m_c$
\beq
m_c^{\rm pole} = 1.4 \:\:\mbox{GeV}
\label{mcpole}
\eeq
which is in the range $1.2 - 1.9$ GeV quoted in \cite{PDG}.
These are the quark threshold values we use in evolving
the Wilson coefficients down to the 1 GeV scale. We have checked
that varying $m_b^{\rm pole}$ 
within the $4.5 - 4.9$ GeV range affects the final values
of the Wilson coefficients at the 0.1\% level, while varying the
charm pole mass 
between 1.2 and 1.9 GeV affects the results at the 15\% level at most.

Even though not all the operators in \eq{Q1-10} are independent, this basis
is of particular interest for 
the present numerical analysis because it is that employed
for the calculation of the Wilson coefficients 
to the NLO order in $\alpha_s$ and $\alpha_e$~\cite{monaco,roma}.

In tables 1 and 2  we give explicitly 
the Wilson coefficients $y_i(\mu)$ of the ten
operators at the scale $\mu = 1.0$ and 0.8 GeV, respectively, in the 
HV and NDR schemes. In the $\chi$QM the chiral symmetry breaking scale
$\Lambda_\chi$ turns out to be about 0.8 GeV. 
This sets a preferential scale
for the matching of the hadronic matrix elements to the Wilson coefficients.
We have checked that the QCD perturbative expansion is under control.
In fact the difference between LO and NLO results for all physical amplitudes
considered---both real and imaginary parts---at $\mu=0.8$ GeV
remains always smaller than 30\%. 
\begin{table}
\begin{small}
\begin{center}
\begin{tabular}{|c|r r||r r||r r|}
\hline
$\Lambda_{QCD}^{(4)}$ & \multicolumn{2}{c||}{ 250 MeV }
                      & \multicolumn{2}{c||}{ 350 MeV } 
                      & \multicolumn{2}{c| }{ 450 MeV } \\
\hline
$\alpha_s(m_Z)_{\overline{\rm MS}}$ 
                      & \multicolumn{2}{c||}{ 0.113 }
                      & \multicolumn{2}{c||}{ 0.119 } 
                      & \multicolumn{2}{c| }{ 0.125 } \\
\hline 
\multicolumn{7}{c}{\mbox{HV}}\\
\hline 
$y_3$&$(0.0007)$&$0.0301$&$(0.0007)$&$0.0390$&$(0.0007)$&$0.0509$ \\
\hline
$y_4$&$(0.0011)$&$-0.0513$&$(0.0012)$&$-0.0610$&$(0.0012)$&$-0.0723$ \\
\hline
$y_5$&$(-0.0004)$&$0.0137$&$(-0.0004)$&$0.0163$&$(-0.0004)$&$0.0209$ \\
\hline
$y_6$&$(0.0011)$&$-0.0766$&$(0.0012)$&$-0.103$&$(0.0012)$&$-0.144$ \\
\hline
$y_7/\alpha$&$(0.172)$&$-0.0115$&$(0.172)$&$-0.0103$&$(0.172)$&$-0.0083$ \\
\hline
$y_8/\alpha$&$(0)$&$0.167$&$(0)$&$0.230$&$(0)$&$0.328$ \\
\hline
$y_9/\alpha$&$(-1.19)$&$-1.71$&$(-1.19)$&$-1.83$&$(-1.19)$&$-2.00$ \\
\hline
$y_{10}/\alpha$&$(0)$&$0.750$&$(0)$&$0.859$&$(0)$&$1.10$ \\
\hline
\multicolumn{7}{c}{\mbox{NDR}}\\
\hline
$y_3$&$(0.0017)$&$0.0268$&$(0.0018)$&$0.0336$&$(0.0018)$&$0.0416$ \\
\hline
$y_4$&$(-0.0019)$&$-0.0491$&$(-0.0021)$&$-0.0574$&$(-0.0022)$&$-0.0660$ \\
\hline
$y_5$&$(0.0007)$&$0.0031$&$(0.0007)$&$-0.0028$&$(0.0007)$&$-0.0165$ \\
\hline
$y_6$&$(-0.0019)$&$-0.0849$&$(-0.0021)$&$-0.119$&$(-0.0022)$&$-0.178$ \\
\hline
$y_7/\alpha$&$(0.149)$&$-0.0119$&$(0.149)$&$-0.0118$&$(0.136)$&$-0.0127$ \\
\hline
$y_8/\alpha$&$(0)$&$0.153$&$(0)$&$0.212$&$(0)$&$0.304$ \\
\hline
$y_9/\alpha$&$(-1.22)$&$-1.71$&$(-1.22)$&$-1.83$&$(-1.22)$&$-1.99$ \\
\hline
$y_{10}/\alpha$&$(0)$&$0.674$&$(0)$&$0.843$&$(0)$&$1.07$ \\
\hline
\end{tabular}
\end{center}
\end{small}
\caption{NLO Wilson coefficients at $\mu=1.0$ GeV for 
$\overline{m}_t(m_W)=183$~GeV, which corresponds to $m_t^{pole}=180$~GeV.
The values of the coefficients at $\mu=m_W$ are given in parenthesis
($\alpha=1/128$).
In addition one has $y_{1,2}(\mu) = 0$.}
\end{table}

In order to test the $\mu$ independence
of $\varepsilon'/\varepsilon$ we vary the matching scale between 0.8 and 1.0
GeV, the highest energy up to which we trust the chiral loop corrections computed 
in I.
We find that, in spite of the fact that some of the Wilson coefficients
vary in this range by up to 50\%, the matching with our matrix elements
reduces the $\mu$-dependence in $\varepsilon'/\varepsilon$ below
20\% in most of the parameter space. We consider this improved stability
a success of the approach.
\begin{table}
\begin{small}
\begin{center}
\begin{tabular}{|c|r r||r r||r r|}
\hline
$\Lambda_{QCD}^{(4)}$ & \multicolumn{2}{c||}{ 250 MeV }
                      & \multicolumn{2}{c||}{ 350 MeV } 
                      & \multicolumn{2}{c| }{ 450 MeV } \\
\hline
$\alpha_s(m_Z)_{\overline{\rm MS}}$ 
                      & \multicolumn{2}{c||}{ 0.113 }
                      & \multicolumn{2}{c||}{ 0.119 } 
                      & \multicolumn{2}{c| }{ 0.125 } \\
\hline 
\multicolumn{7}{c}{\mbox{HV}}\\
\hline 
$y_3$&$(0.0007)$&$0.0338$&$(0.0007)$&$0.0456$&$(0.0007)$&$0.0602$ \\
\hline
$y_4$&$(0.0011)$&$-0.0522$&$(0.0012)$&$-0.0626$&$(0.0012)$&$-0.0741$ \\
\hline
$y_5$&$(-0.0004)$&$0.0140$&$(-0.0004)$&$0.0192$&$(-0.0004)$&$0.0397$ \\
\hline
$y_6$&$(0.0011)$&$-0.0904$&$(0.0012)$&$-0.137$&$(0.0012)$&$-0.242$ \\
\hline
$y_7/\alpha$&$(0.172)$&$-0.0131$&$(0.172)$&$-0.0111$&$(0.172)$&$-0.0039$ \\
\hline
$y_8/\alpha$&$(0)$&$0.209$&$(0)$&$0.327$&$(0)$&$0.583$ \\
\hline
$y_9/\alpha$&$(-1.19)$&$-1.82$&$(-1.19)$&$-2.04$&$(-1.19)$&$-2.45$ \\
\hline
$y_{10}/\alpha$&$(0)$&$0.835$&$(0)$&$1.14$&$(0)$&$1.66$ \\
\hline
\multicolumn{7}{c}{\mbox{NDR}}\\
\hline
$y_3$&$(0.0017)$&$0.0294$&$(0.0018)$&$0.0373$&$(0.0018)$&$0.0422$ \\
\hline
$y_4$&$(-0.0019)$&$-0.0493$&$(-0.0021)$&$-0.0569$&$(-0.0022)$&$-0.0603$ \\
\hline
$y_5$&$(0.0007)$&$-0.0014$&$(0.0007)$&$-0.0167$&$(0.0007)$&$-0.0708$ \\
\hline
$y_6$&$(-0.0019)$&$-0.104$&$(-0.0021)$&$-0.171$&$(-0.0022)$&$-0.353$ \\
\hline
$y_7/\alpha$&$(0.149)$&$-0.0138$&$(0.149)$&$-0.0156$&$(0.149)$&$-0.0274$ \\
\hline
$y_8/\alpha$&$(0)$&$0.189$&$(0)$&$0.294$&$(0)$&$0.511$ \\
\hline
$y_9/\alpha$&$(-1.22)$&$-1.81$&$(-1.22)$&$-2.03$&$(-1.22)$&$-2.42$ \\
\hline
$y_{10}/\alpha$&$(0)$&$0.819$&$(0)$&$1.11$&$(0)$&$1.60$ \\
\hline
\end{tabular}
\end{center}
\end{small}
\caption{Same as in Table 1 at $\mu=0.8$ GeV.}
\end{table}
\clearpage

\section{The Hadronic Matrix Elements}

In paper I we have computed all hadronic matrix elements of the effective
quark operators in \eq{Q1-10} in the framework of the $\chi$QM. The matrix 
elements are obtained by the integration of the constituent quarks by means of
dimensional regularization. The loop
integration leads to results that  depend on the scheme employed to 
deal with $\gamma_5$ but are scale independent. The renormalization-scale 
dependence
is introduced in our approach by the meson-loop renormalization of the
amplitudes, as explained in I. 
The meson-loop corrections together with the gluon-condensate 
contributions are the most relevant ingredients in reproducing
the $\Delta I = 1/2$ selection rule in $K\to \pi\pi$ decays, (as discussed
in II). 

The $\chi$QM results are expressed in a double power expansion on 
$M^2/\Lambda_{\chi}^2$ and $p^2/\Lambda_{\chi}^2$, 
where $M$ is a dimensionful parameter of the
model which is not determined (generically, it can be
interpreted as the constituent quark mass in mesons) and $p$
is a typical external momentum. 

The value of $M$ is
constrained~\cite{B} by experimental data on the decay of $\pi^0$ and $\eta$ to
be
\beq
M = 223 \pm 9 \: \mbox{MeV}
\eeq
(and $M = 243 \pm 9$ MeV if higher order corrections are included). The value of
\beq
M = m_{\rho}/\sqrt{12} \simeq 222\ \mbox{MeV} 
\eeq
is found by vector-meson-dominance estimates. Finally, in a recent fit of all
input parameteres of the extended Nambu-Jona-Lasinio model~\cite{B2},
 it was found a value of
\beq
M \simeq 200 \: \mbox{MeV} \, .
\eeq

While we could simply take these values and thus make the $\chi$QM predictive,
our approach also allows for a self-consistent determination of a 
range for $M$ that can be compared to the above
values. 

The idea is that in physical observables
the $\gamma_5$-scheme and $\mu$-dependences of the
matrix elements  should balance the corresponding dependences
of the NLO Wilson coefficients . 
In I we have constructed the complete $O(p^2)$ chiral representation
of the lagrangian in \eq{ham}, where the local quark operator
$Q_i$ is represented by a linear combination of bosonic operators
$B_\alpha$, namely $Q_i \to \sum_\alpha G_\alpha(Q_i) B_\alpha$.
The effective quark lagrangian is therefore replaced
by the following chiral representation
\beq
{\cal L}^{\Delta S = 1}_\chi = 
-\frac{G_F}{\sqrt{2}}\ \sum_{i,\alpha}\ C_i (\mu)\ 
           G_\alpha (Q_i)\ B_\alpha  
\, . 
\label{chiham}
\eeq

As mentioned above, and 
discussed at lenght in I, the chiral coefficients  
$G_\alpha$ determined 
via the $\chi$QM approach are 
$\gamma_5$-scheme dependent.
While the $\gamma_5$-scheme dependence arises in the $\chi$QM
from the integration of the chiral fermions, the explicit 
$\mu$-dependence is entirely due to the chiral loop
renormalization of the matrix elements:
\beq
\vev{b|{\cal L}^{\Delta S = 1}_\chi|a} = 
-\frac{G_F}{\sqrt{2}}\ \sum_{i,\alpha}\ C_i (\mu_{SD})\ 
           G_\alpha (Q_i)\ \vev{b|B_\alpha|a} (\mu_{LD}) 
\, , 
\label{chimatr}
\eeq
where we have labeled by $a$ and $b$ the initial and final bosonic states.
We remark that in our approach the $\mu$-dependence of the
chiral loops is not cancelled by higher order counterterms, as it
is usually required in the strong chiral lagrangian. 

The renormalization scale 
dependence is therefore determined order by order in the
energy expansion of the chiral lagrangian. 
In this respect there is no direct counterpart to the expansion
in strong and electromagnetic
couplings on which the short-distance
analysis is based and, accordingly, we refer to the
explicit $\mu$-dependence in the matrix elements as to
the long-distance (LD) or ``non-perturbative'' scale dependence.
A purely perturbative renormalization scale dependence is introduced
in the matrix elements by the NLO running of the quark condensate,
which we include whenever a comparison between values at
different scales is required. Otherwise, quark and gluon condensates
are considered in our approach as phenomenological parameters. 

Our aim is to test whether the estimate of observables
is consistently improved by matching the ``long-distance'' 
$\gamma_5$-scheme and $\mu$ dependences so obtained
with those present in the short-distance analysis (in particular
we identify $\mu_{SD}$ with $\mu_{LD}$).
Whether and to what extent
such an improvement is reproduced for many observables and
for a consistent set of parameters, might tell us how well low-energy
QCD is modelled in the $\chi$QM-chiral lagrangian approach  that
we have devised.

In II, we have shown that minimizing the $\gamma_5$-scheme dependence
of the physical 
isospin $I=0$ and 2 amplitudes determines a range for the parameter
$M$  between 160 and 220 MeV. 
In II, it  was also found that the $\mu$ dependence induced
by the Wilson coefficients is substantially reduced by that of the
hadronic matrix elements. 

These issues become crucial for $\varepsilon'/\varepsilon$ where
the $\gamma_5$-scheme dependence induced by the Wilson coefficients determines
an uncertainty as large as 80\% when using
the $1/N_c$ hadronic matrix elements (see for instance ref.~\cite{monaco})
which are scheme independent.

In the following, for the reader's convenience,
we report from I the expressions for the isospin amplitudes for all ten
operators in \eq{Q1-10}:
\beq
\langle Q_i \rangle_{0,2} \equiv \langle 2 \pi, I =0,2 | Q_i | K^0 \rangle \,  .
\eeq
The corresponding 
one-loop meson corrections are denoted by $a_{0,2}(Q_i)$.
The Clebsh-Gordan coefficients for the isospin projections 
can be found in I.

For the HV case we obtain:
\bea
\langle Q_1 \rangle _0 & = & \frac{1}{3} X \left[ -1 + \frac{2}{N_c} \left(
1 - \delta_{\vev{GG}} \right)
\right] + a_0 (Q_1)\\
\langle Q_1 \rangle _2 & = & \frac{\sqrt{2}}{3} X \left[ 1 + \frac{1}{N_c}
\left(
1 - \delta_{\vev{GG}} \right)
\right] + a_2 (Q_1)\\
\langle Q_2 \rangle _0 & = & \frac{1}{3} X \left[ 2  - \frac{1}{N_c} \left(
1 - \delta_{\vev{GG}} \right)
\right] + a_0 (Q_2)\\
\langle Q_2 \rangle _2 & = &  \frac{\sqrt{2}}{3} X \left[ 1 + \frac{1}{N_c}
\left( 1 - \delta_{\vev{GG}} \right) \right] + a_2 (Q_2)\\
\langle Q_3 \rangle _0 & = & \frac{1}{N_c} X  \left(
1 - \delta_{\vev{GG}} \right) + a_0 (Q_3)\\
\langle Q_4 \rangle _0 & = & X + a_0 (Q_4) \\
\langle Q_5 \rangle _0 & = &  \frac{2}{N_c}  \, 
\frac{\langle \bar{q}q \rangle}{M f_\pi^2} \, X'  + a_0 (Q_5)\\
\langle Q_6 \rangle _0 & = &  2 \, 
\frac{\langle \bar{q}q \rangle}{M f_\pi^2} \, X' + a_0 (Q_6) \\
\langle Q_{7} \rangle _{0} & = & \frac{2 \sqrt{3}}{N_c} \, 
\frac{\langle \bar{q} q \rangle ^2}{f_\pi^3} 
- \frac{1}{N_c} \, 
\frac{\langle \bar{q}q \rangle}{M f_\pi^2} \, X'
  \nnu \\
 & & - \frac{2}{N_c} \frac{\vev{\bar{q}q}}{M f_\pi^2} \ Y
 + \frac{1}{2} X  + a_{0}(Q_7)\\
 \langle Q_{7} \rangle _{2} & = & \sqrt{6} \,
 \frac{\langle \bar{q} q \rangle ^2}{f_\pi^3} \frac{1}{N_c}
  - \frac{\sqrt{2}}{N_c} \frac{\vev{\bar{q}q}}{M f_\pi^2} \ Y
  - \frac{\sqrt{2}}{2} X + a_{2}(Q_7)\\
\langle Q_8 \rangle _0 & = &  2 \, \sqrt{3} \, 
\frac{\langle \bar{q} q \rangle ^2}{f_\pi^3} - 
\frac{\langle \bar{q}q \rangle}{M f_\pi^2} \, X' 
 \nnu \\
& & - 2 \, \frac{\vev{\bar{q}q}}{M f_\pi^2} \ Y
    + \frac{1}{2 N_c} X  \left(1 + \delta_{\vev{GG}}\right) + a_0(Q_8)\\
\langle Q_8 \rangle _2 & = &
\sqrt{6} \, \frac{\langle \bar{q} q \rangle ^2}{f_\pi^3}
  - \sqrt{2} \frac{\vev{\bar{q}q}}{M f_\pi^2} \ Y
  - \frac{\sqrt{2}}{2 N_c} X \left(1 + \delta_{\vev{GG}}\right)   + a_2(Q_8) \\
\langle Q_9 \rangle _0 & = &  - \frac{1}{2} X \left[ 1 - \frac{1}{N_c}
\left( 1 - \delta_{\vev{GG}} \right) \right] + a_0 (Q_9) \\
\langle Q_9 \rangle _2 & = &   \frac{\sqrt{2}}{2} X \left[ 1 + \frac{1}{N_c}
\left( 1 - \delta_{\vev{GG}} \right) \right] + a_2 (Q_9)\\
\langle Q_{10} \rangle _0 & = &   \frac{1}{2} X \left[ 1 - \frac{1}{N_c}
\left( 1 - \delta_{\vev{GG}} \right) \right] + a_0(Q_{10}) \\
\langle Q_{10} \rangle _2 & = &   \frac{\sqrt{2}}{2} X \left[ 1 + \frac{1}{N_c}
\left( 1 - \delta_{\vev{GG}} \right) \right] + a_2 (Q_{10})\, .
 \eea
where
\beq
X \equiv \sqrt{3} f_\pi \left( m_K^2 - m_\pi^2 \right)\ , 
\quad\quad 
X'  =   X \left( 1 - 6\,
\frac{M^2}{\Lambda_\chi^2} \right)
\eeq
and
\beq
Y \equiv \sqrt{3} f_\pi \left[m_\pi^2 + 3\ m_K^2
\frac{M^2}{\Lambda^2_\chi}\right] \ ;
\eeq
$\delta_{\vev{GG}}$ is given by 
\beq
 \delta_{\langle GG \rangle} = \frac{N_c}{2} \frac{\langle
 \alpha_s G G/\pi \rangle}{16 \pi^2 f^4} \, . \label{GG}
\eeq
It is (\ref{GG}) that parameterizes the non-perturbative part of the computation
by the contribution of the gluon condensate
$\langle
 \alpha_s G G/\pi \rangle$, as discussed in I.
 
The 
renormalization of $f$ is taken into account by replacing
$f$ with the one-loop parameter $f_1$ in
the tree-level amplitudes, which amounts to replacing $1/f^3$ with 
$1/f^3_\pi$
multiplied by
\beq
1 + 3\ \frac{f_\pi - f_1}{f_\pi} \simeq 1.18\, .
\eeq

In the NDR case we  find:
 \bea
\langle Q_1 \rangle _0 & = & \frac{1}{3} X \left[ -1 + \frac{2}{N_c} \left(
1 - \delta_{\vev{GG}} \right)
\right] + a_0 (Q_1)\\
\langle Q_1 \rangle _2 & = & \frac{\sqrt{2}}{3} X \left[ 1 + \frac{1}{N_c}
\left(
1 - \delta_{\vev{GG}} \right)
\right] + a_2 (Q_1)\\
\langle Q_2 \rangle _0 & = & \frac{1}{3} X \left[ 2  - \frac{1}{N_c} \left(
1 - \delta_{\vev{GG}} \right)
\right] + a_0 (Q_2)\\
\langle Q_2 \rangle _2 & = &  \frac{\sqrt{2}}{3} X \left[ 1 + \frac{1}{N_c}
\left( 1 - \delta_{\vev{GG}} \right) \right] + a_2 (Q_2)\\
\langle Q_3 \rangle _0 & = & \frac{1}{N_c} \left( X'
-  \delta_{\vev{GG}} X \right) + a_0 (Q_3)
\\
\langle Q_4 \rangle _0 & = & X'  +a_0 (Q_4)\\
\langle Q_5 \rangle _0 & = & \frac{2}{N_c}  \,
\frac{\langle \bar{q}q \rangle}{M f_\pi^2} \, X''  + a_0 (Q_5)  \\
\langle Q_6 \rangle _0 & = & 2  \,
\frac{\langle \bar{q}q \rangle}{M f_\pi^2} \, X''  + a_0 (Q_6)  \\
\langle Q_{7} \rangle _{0} & = &  \frac{2 \sqrt{3}}{N_c} \, 
\frac{\langle \bar{q} q \rangle ^2}{f_\pi^3}
 \left( 1 - 3 \frac{M^3 f_\pi^2}{\langle \bar{q} q \rangle \Lambda_{\chi}^2}
 \right) - \frac{1}{N_c} \frac{\langle \bar{q}q \rangle}{M f_\pi^2} \, X'' 
  \nnu \\
 & & - \frac{2}{N_c} \frac{\vev{\bar{q}q}}{M f_\pi^2} \ Y'
 + \frac{1}{2} X + a_{0} (Q_7)\\
\langle Q_{7} \rangle _{2} & = &
\frac{1}{N_c} \sqrt{6} \, \frac{\langle \bar{q} q \rangle ^2}{f_\pi^3}
\left(1 - 3 \frac{M^3 f_\pi^2}{\langle \bar{q} q \rangle \Lambda_{\chi}^2}
\right)  \nnu \\
& &  - \frac{\sqrt{2}}{N_c} \frac{\vev{\bar{q}q}}{M f_\pi^2} \ Y'
- \frac{\sqrt{2}}{2} X + a_{2} (Q_7)\\
\langle Q_8 \rangle _0 & = &   2 \sqrt{3}\,
 \frac{\langle \bar{q} q \rangle ^2}{f_\pi^3}
 \left( 1 - 3 \frac{M^3 f_\pi^2}{\langle \bar{q} q \rangle \Lambda_{\chi}^2}
 \right) - \frac{\langle \bar{q}q \rangle}{M f_\pi^2} \, X'' 
 \nnu \\
& &  - 2 \, \frac{\vev{\bar{q}q}}{M f_\pi^2} \ Y' \:
+ \frac{1}{2 N_c} X \left(1 + \delta_{\vev{GG}}\right) + \: a_0(Q_8) \\
\langle Q_8 \rangle _2 & = &
\sqrt{6} \, \frac{\langle \bar{q} q \rangle ^2}{f_\pi^3}
\left(1 - 3 \frac{M^3 f_\pi^2}{\langle \bar{q} q \rangle
\Lambda_{\chi}^2} \right) \nnu \\
& &  - \sqrt{2} \frac{\vev{\bar{q}q}}{M f_\pi^2} \ Y'
 - \frac{\sqrt{2}}{2 N_c} X \left(1 + \delta_{\vev{GG}}\right)  \: +
a_2(Q_8)\\
\langle Q_9 \rangle _0 & = &  - \frac{1}{2} \left[ X - \frac{1}{N_c}
\left(2 X - X' - \delta_{\vev{GG}} X\right) \right] + a_0 (Q_9) \\
\langle Q_9 \rangle _2 & = &   \frac{\sqrt{2}}{2} X \left[ 1 + \frac{1}{N_c}
\left( 1 - \delta_{\vev{GG}} \right) \right] + a_2 (Q_9)\\
\langle Q_{10} \rangle _0 & = &   \frac{1}{2} \left[2 X - X' - \frac{1}{N_c}
\left(1 - \delta_{\vev{GG}}\right) X \right] + a_0(Q_{10}) \\
\langle Q_{10} \rangle _2 & = &   \frac{\sqrt{2}}{2} X \left[ 1 + \frac{1}{N_c}
\left( 1 - \delta_{\vev{GG}} \right) \right] + a_2 (Q_{10})\, .
\eea
where
\beq
X'' = X \left( 1 - 9\ \frac{M^2}{\Lambda_\chi^2} \right) \, ,
\quad Y' \equiv \sqrt{3} f_\pi \left[m_\pi^2 + 3\ 
\left(m_K^2-m_\pi^2\right)
\frac{M^2}{\Lambda^2_\chi}\right]
\ .
\eeq

$\langle Q_i \rangle _2 = 0$ for $i = 3,4,5,6$ in both schemes.

Of particular interest  are the matrix elements $\langle Q_{6} \rangle _0 $
and $\langle Q_{8} \rangle _2 $ which dominate any estimate
of $\varepsilon'/\varepsilon$; their leading
effect  was included in the toy model of ref.~\cite{BEF}.

The most striking feature concerning the gluon-penguin operators is
the linear dependence on the quark condensate that is found in
the $\chi$QM in
 contrast to the quadratic one of the VSA. This difference explains 
 the different weight that these operators have in the two models.
 
Concerning the electroweak-penguin operators, as discussed in I,
the terms proportional 
to $Y$---so far
neglected in all estimates---give an important contribution
that makes  the electroweak-penguin operators larger and, 
accordingly,
the cancellation between electroweak and gluon penguins   
effective even for the present
values of $m_t$.
We shall come back to this point in  section 5.

\section{The Input Parameters}

 The quark and the gluon condensates are two  input parameters of
 our computation. As discussed in I,
 their phenomenological determination  
is a complicated 
question (they parameterize the genuine non-perturbative part of the 
computation) and the literature offers different estimates. 

For guidance,
we identify the condensates entering our computation with those obtained by
fitting the experimental data  by means of the QCD sum 
rules (QCD--SR) or
lattice computations.

A review of recent determinations of these parameters, together
 with a justification of the range below, is given in I. 
Here we only report the ranges that we will explore in our
numerical analysis.

For the gluon condensate, we take 
the scale independent range
\beq
\langle \frac{\alpha_s}{\pi} G G \rangle = (376 \pm  47 \:
\mbox{MeV} )^4 \, , \label{GGexp}
\eeq
which encompasses the results of recent QCD-SR
analysis~\cite{Narison}. While this is a crucial input parameter in the physics
of the $\Delta I =1/2$ rule (see II), it plays only a minor role
in a penguin-dominated quantity like $\varepsilon '/\varepsilon$.

For the quark condensate, we consider the range
\beq
 - ( 200 \: \mbox{MeV} )^3 \leq 
 \vev{\bar{q}q} \leq  - ( 280 \: \mbox{MeV} )^3 
\label{qqexp}
\eeq 
which includes the central values and the errors
of the  QCD-SR~\cite{DdeR} and lattice estimates~\cite{qqlattice1}.

The rather conservative range (\ref{qqexp}) is the one advocated in I. In
 II it is shown that the $\Delta I =1/2$ selection rule seems to prefer the upper
 half of this range.

 For comparison, in most
estimates of $\varepsilon'/\varepsilon$ the PCAC value 
\beq
\vev{\bar{q}q} (\mu) = - \frac{f_K^2 m_K^2 ( 1 - 
\delta_K)}{\overline{m}_s(\mu) + \overline{m}_d (\mu)} \, , \label{qqQCD}
\eeq
is taken (with $\delta_K$ equal to zero) and the error range is that of 
the determination of $m_s$~\cite{JaminAlton}:
\beq
\overline{m}_s( 1 \: {\rm GeV}) = 178 \pm 18 \: {\rm MeV} \, .
\eeq 
This choice gives a quark condensate of
\beq
 \vev{\bar{q}q} =  - ( 261 \pm 9 \: \mbox{MeV} )^3
 \eeq
 at $\mu = 1.0$ GeV that corresponds, via NLO renormalization at fixed 
 $\Lambda_{\rm QCD}^{(4)} = 350$ MeV, to
\beq
 \vev{\bar{q}q} =  - ( 244 \pm 9 \: \mbox{MeV} )^3 \label{rangeII} 
 \eeq
 at the matching scale of $\mu = 0.8$ GeV. The quark condensate in
 (\ref{rangeII}) has
an uncertainty of  only 9 MeV, that is perhaps
 too small a range to
 account for the actual uncertainty.
 For instance,
 if the QCD-SR improved estimate~\cite{deltaK}
\beq
\delta_K = 0.34 ^{+0.23}_{-0.17} \label{dK}
\eeq
is taken into account, the  
 range (at $\mu = 1$ GeV) to
be explored becomes comparable to that of (\ref{qqexp}):
\beq
 - ( 183 \: \mbox{MeV} )^3 \leq 
 \vev{\bar{q}q} \leq  - ( 256 \: \mbox{MeV} )^3 \label{rangeIII} \, ,
\eeq
with a central value  much smaller than in (\ref{rangeII}). 

The range
(\ref{rangeIII}) is consistent with what one finds by the
same PCAC value as in (\ref{qqQCD}) but with $f_K$ and $m_K$ replaced by,
respectively, $f_\pi$ and $m_\pi$, and 
$\overline{m}_s(\mu) + \overline{m}_d (\mu)$ by
\beq
\overline{m}_u(1 \: {\rm GeV}) + 
\overline{m}_d (1 \: {\rm GeV}) = 12 \pm 2.5 \: {\rm MeV} \, ,
\eeq
as given in ~\cite{BPdeR}. By means of the latter,
 taken at $\mu = 0.8$ GeV, we find
\beq
\vev{\bar{q}q} = 
\left( - 221 \pm 17 \:\:\:{\rm MeV} \right) ^3 \label{rangeIV} \, .
\eeq
The range (\ref{rangeIV}) suffers of a larger error with respect to that
of (\ref{rangeII}), which however does not take into account
(\ref{dK}) and, accordingly, the  much broader range (\ref{rangeIII}) which is
more realistic.

Because of such uncertainties,
we will consider in section 7 two possible ranges for the quark condensate:
 the  range (\ref{qqexp}) for our  most conservative
 estimate and the improved PCAC result  (\ref{rangeIV}) for a second,
 more restrictive one.  
As discussed in the introduction, these two ranges complete
each other in providing, at the same time,
a definite prediction and a gauge of the overall uncertainty of the prediction
itself.

In section 7, 
in order to make the comparison to other estimates easier,
 we will also give our result
for the range 
(\ref{rangeII}).

\section{The $B_i$ Factors}

Let us introduce the effective factors
\beq
B_i^{(0,2)} \equiv \frac{\langle Q_i \rangle _{0,2}^{\chi {\rm QM}}}
{\langle Q_i \rangle _{0,2}^{\rm VSA}} \, ,
\eeq
which give the ratios between our hadronic matrix elements and those of the
VSA. They are a useful way of comparing different evaluations.

In table 3, we collect the $B_i$ factors for the ten
operators. The values of the $B_i$ depend on the
scale at which the matrix elements
 are evaluated,  the input parameters and
$M$; moreover, in the $\chi$QM they depend on the
$\gamma_5$-scheme employed.
We have given in table 3 a representative 
example of their values and variations.

The values of $B_1^{(0)}$ and $B_2^{(0)}$ show 
that the corresponding
hadronic matrix elements in the $\chi$QM are, once non-factorizable
contributions and meson renormalization have been included, respectively
about ten and three times
larger than their VSA values. 
At the same time, 
$B_1^{(2)}$ and $B_2^{(2)}$ turn out to be 
at most half of what found in the VSA (for the starred entries see the 
comment at the end of the section). 
These features make
it possible for the selection rule to be reproduced in the $\chi$QM, as extensively
discussed in II.

For comparison, in the $1/N_c$ approach of ref. \cite{1/N}, the inclusion of
meson-loop renormalization through a cutoff regularization, leads, at the
scale of 1 GeV, to $B_1^{(0)} = 5.2$, $B_2^{(0)} = 2.2$ and 
$B_1^{(2)} = B_2^{(2)} = 0.55$, a result that is not sufficient to
reproduce the $\Delta I = 1/2$ rule.
The similarity of the HV values $B_1^{(2)} = B_2^{(2)} = 0.55$ obtained in
the $\chi$QM with the corresponding $1/N_c$ results 
is remarkable, and yet
a numerical coincidence, since the suppression originates from gluon
condensate corrections in the $\chi$QM, whereas it is the effect of
the meson loop renormalization (regularized via explicit cut-off)
in the analysis of ref. \cite{1/N}. 

The values of the penguin matrix elements $\vev{Q_3}$ 
and $\vev{Q_4}$ in the $\chi$QM lead to rather large $B_i$ factors. 
In the case of $Q_3$, the $\chi$QM result has the opposite sign of the VSA
result and $B_3$ is negative.
This is the effect of the large non-perturbative gluon
correction. 

Regarding the gluon penguin operator $Q_6$ (and $Q_5$), we find that 
the $\chi$QM gives
a result consistent with the VSA (and the $1/N_c$ approach), 
$B_6$ ($B_5$) being approximately equal to two for
small values of the quark condensate and one-half at larger values. It is the
quadratic dependence (to be contrasted to the linear dependence in the
$\chi$QM) of the VSA matrix element for the penguin operators that
it responsible for the different weight of these operators at different values
of the quark condensate. 
\begin{table}
\begin{small}
\begin{center}
\begin{tabular}{|c||c|c|c|c|}
\hline
 & \multicolumn{2}{|c|}{\rm HV} & \multicolumn{2}{c|}{\rm NDR} \\ 
\cline{2-5}
\cline{2-5}
 & $\mu = 0.8$ GeV & $\mu = 1.0$ GeV & $\mu = 0.8$ GeV & $\mu = 1.0$ GeV\\
\hline
$B^{(0)}_1$  & 10.6 & 11.1 & 10.6 & 11.1\\
\hline
$B^{(0)}_2$  & 2.8 & 3.0  & 2.8 & 3.0\\
\hline
$B^{(2)}_1$  & 0.52 & 0.55 &  0.52 & 0.55 \\
\hline
$B^{(2)}_2$ & 0.52 & 0.55 & 0.52 & 0.55 \\
\hline
$B_3$ & $-2.9$ & $-3.0$ & $-3.7$ & $-3.9$\\
\hline
$B_4$ & 1.8 & 1.9 & 1.0 & 1.1\\
\hline
$B_5 = B_6$ & $1.7 \div 0.61$ & $1.8 \div 0.64$ & $1.0 \div 0.38$ & $1.1 \div 0.40$\\
\hline
$B_7^{(0)}$ &$3.0 \div 2.2$ &  $3.3 \div 2.4$ & $2.9 \div 2.2$  & $3.2 \div 2.3$ \\
\hline
$B_8^{(0)}$ &$3.3 \div 2.2$& $3.6 \div 2.4$ &  $3.2 \div 2.2$ & $3.5 \div 2.4$ \\
\hline
$B_9^{(0)}$ &3.9 & 4.0 & 3.5  & 3.6\\
\hline
$B_{10}^{(0)}$ &4.4 & 4.7 & 5.6 & 5.9\\
\hline
$B_7^{(2)}$ &$2.7 \div 1.5$ & $3.0 \div 1.5$ & $2.7 \div 1.4$ & $2.9 \div 1.5$\\
\hline
$B_8^{(2)}$ &$2.1 \div 1.4 $& $2.3 \div 1.5$ & $2.1 \div 1.4$ & $2.3 \div 1.5$\\
\hline
$B_9^{(2)}$ &0.52 & 0.55 & 0.52 & 0.55 \\
\hline
$B_{10}^{(2)}$ & 0.52 & 0.55 & 0.52 & 0.55 \\
\hline
\end{tabular}
\end{center}
\end{small}
\caption{The $B_i$ factors in the $\chi$QM (including meson-loop
renormalizations) at two different scales: $\mu$ = 0.8 and 1.0 GeV and in the
two $\gamma_5$-schemes. We have taken
the gluon
condensate at the central value of \eq{GGexp}, while the range given
for $B_{5-8}$
corresponds to varying the quark condensate according to \eq{qqexp}. The 
results shown are given  for $M = 220$ MeV.}
\end{table}
The lattice estimate at $\mu = 2$ GeV 
for these operators 
gives $B_5 = B_6 = 1.0 \pm 0.2$~\cite{martinelli}.
A direct comparison in this case is not possible.

The electroweak $B_i$ factors are all larger in the $\chi$QM than in the VSA,
except for $B_{9,10}^{(2)}$ that are about
 1/2 in the HV and about 0.4 in the NDR scheme. 
For comparison, 
the lattice estimate at $\mu = 2$ GeV
in this case yields $B_{7,8}^{(2)} = 1.0 \pm 0.2$ and
$B_9^{(2)} = 0.62 \pm 0.10$~\cite{martinelli}.

The most relevant result for
$\varepsilon'/\varepsilon$ is the value of $B_8^{(2)}$ which ranges
from 1.5 to 2 times that of the VSA. This increase is due to two independent
reasons. On the one hand, we found two new terms 
in the chiral lagrangian that
have not been included so far in the VSA estimate of the matrix elements. 
The chiral coefficients of these terms are 
computed in the $\chi$QM approach---as
discussed in detail in I. 
It is an open question how they can be determined in the VSA framework.

From this point of view,
what we have referred to as VSA---and used in table 3 as normalization for the 
$Q_{7,8}$ operators---is not the complete VSA result. 
The inclusion of the new terms amounts up to a 60\%
increase of $B_{7,8}^{(2)}$ for small values of $\vev{\bar qq}$ 
in the chosen range and down to about 10\%
for large values; smaller effects are found in the case of  $B_{7,8}^{(0)}$. 
On the other hand, the meson-loop renormalization 
associated with the new chiral terms is large (see
I) and adds up to reproduce the results shown in table 3.
The increase in importance of the operator $Q_8$ with
respect to $Q_6$ turns into a
 more effective cancellation between the two operators for large
values of the quark condensate while at smaller
values the gluon penguin contribution prevails. 

The relations $B_9^{(2)} = B_2^{(2)}$, $B_{10}^{(2)} = B_1^{(2)}$ and
$B_1^{(2)} = B_2^{(2)}$
hold true in both $\gamma_5$-schemes. 
These relationships are a reminiscent of those among
the operators---which are preserved by gluon corrections and meson 
renormalization.

\clearpage

\section{Studying $\varepsilon'/\varepsilon$ in the $\chi$QM}

The quantity
 $\varepsilon'/\varepsilon$ can be written as
\beq
\frac{\varepsilon '}{\varepsilon} =  
\frac{G_F \omega}{2\mod{\epsilon}\Re{A_0}} \:
\mbox{Im}\, \lambda_t \: \:
 \left[ \Pi_0 - \frac{1}{\omega} \: \Pi_2 \right] \, ,
\label{eps}
 \eeq
where, referring to the $\Delta S=1$ quark lagrangian of \eq{ham},
\bea
 \Pi_0 & = &  \sum_i y_i \, \langle  Q_i  \rangle _0 \\
 \Pi_2 & = &  \sum_i y_i \, \langle Q_i \rangle_2  + 
\omega \: \sum_i y_i \, \langle  Q_i  \rangle _0  \: \Omega_{\eta +\eta'} \ ,
\eea
and
\beq
\mbox{Im}\, \lambda_t \equiv \Im V_{td}V^*_{ts} \, .
\eeq

The quantity $\Omega_{\eta +\eta'}$ includes the effect of the isospin-breaking 
mixing between $\pi^0$ and the etas.

Since Im $ \lambda_u =0$ according to the standard conventions,
 the short-distance component of $\varepsilon'/\varepsilon$
is determined by the Wilson coefficients $y_i$. 
Following the approach of
ref.~\cite{monaco}, $y_1(\mu)=$ $y_2(\mu)=0$. 
As a consequence, the matrix elements of $Q_{1,2}$ do not directly
enter the determination of $\varepsilon'/\varepsilon$.
On the other hand, in the HV scheme the matrix elements of $Q_4$
$Q_9$ and $Q_{10}$
can be expressed in terms of those of $Q_{1,2}$ and $Q_3$. 
The work of ref.~\cite{monaco} has taken advantage of this fact
to determine some of the penguin matrix elements, after imposing 
the $\Delta I=1/2$ rule.
The $\chi$QM determination of $\vev{Q_{1,2,3}}_{0}$ gives for 
$\vev{Q_{4}}_0$
a result that differs substantially 
from that used in ref.~\cite{monaco}, as we 
discuss at the end of section 6.3.

We take, as input values for
the relevant quantities, the central values given in appendix. 
We thus have
\beq
\frac{G_F \omega}{2\mod{\epsilon}\Re{A_0}} \simeq 349 \ \mbox{GeV}^{-3},
\qquad 
\omega = 1/22.2\ , \qquad \Omega_{\eta + \eta'} = 0.25 \ .
\label{values}
\eeq

The large value in \eq{values} for $1/\omega$ comes from
the $\Delta I =1/2$ selection rule. In II we have shown that such a rule is
well reproduced by the $\chi$QM evaluation of the hadronic matrix elements. 
As the precise values of $\Re A_0$ and $\omega$  
depend on the  choice of the input parameters and of 
$M$---the selection 
rule being satisfied within a 20\% approximation---we have taken
the corresponding experimental values.
Similarly, the value taken for $\varepsilon$ is the experimental one.

\subsection{The Mixing Parameter $\mbox{Im}\, \lambda_t$}

A range for Im $\lambda_t$ is determined from the experimental value of
$\varepsilon$ as a function of $m_t$ and the other
relevant parameters involved in the theoretical estimate.
We will use the most recent NLO results for the QCD correction factors
$\eta_{1,2,3}$ which are given in the NDR scheme \cite{etas} 
and vary the $\Delta S = 2$ hadronic parameter $\hat B_K$ 
around the central value obtained in the $\chi$QM using the same 
regularization scheme. 

In order to restrict the allowed values of Im $\lambda_t$ we have solved
 the two equations
\beq
\varepsilon_{th} 
 ( \hat{B}_K, |V_{cb}|, |V_{us}|, 
  |V_{ub}|/|V_{cb}|, \Lambda_{\rm QCD}, m_t, m_c, \eta,\rho ) =  
 \varepsilon   
\label{bound1}
\eeq
\beq
\eta^2 + \rho^2 =  \frac{1}{| V_{us}|^2} \frac{|V_{ub}|^2}{|V_{cb}|^2}
 \label{bound2}
\eeq
 to find the allowed values of $\eta$ and $\rho$, given $m_t$, $m_c$
 and~\cite{PDG}  
 \bea
 |\varepsilon| & = & (2.266 \pm 0.023) \times 10^{-3}  \\
 |V_{us}| & = & 0.2205 \pm 0.0018  \\
 |V_{cb}| & = & 0.041 \pm 0.003  \\
 |V_{ub}|/|V_{cb}| & = & 0.08 \pm 0.02 \, . 
\eea
 For the renormalization group invariant parameter
 $\hat B_K$  we take the rather conservative range
 \beq
 \hat B_K = 0.55 \pm 0.25
 \label{BKrange} 
 \eeq
 that 
 encompasses both the $\chi$QM model prediction~\cite{BK} and other current
 determinations~\cite{PP}.
 
 For the NLO order $\eta$-parameters for 
$\Lambda^{(4)}_{\rm QCD} = 350$ MeV and
 $m_t^{(pole)} = 180$ GeV, at $\mu=m_c$ we find
 \beq
 \eta_1 = 1.36\quad \eta_2 = 0.513\quad \eta_3= 0.446
 \eeq
 
 We do not include bounds provided by the quantity $x_d$ 
 of B-physics that we find to 
 have a  marginal impact in the determination of $\varepsilon'/\varepsilon$
 once the large error in (\ref{BKrange}) is taken into account.

 This procedure gives two possible ranges
for $\Im \lambda_t \simeq \eta |V_{us}| |V_{cb}|^2$, 
which correspond to having the KM phase in
the I or II quadrant ($\rho$ positive or negative, respectively). 
For example, for $m_t^{\rm pole} = 180$ GeV 
($\overline{m}_t(m_W) \simeq 183$ GeV) and 
$\Lambda_{\rm QCD}^{(4)}= 350$ MeV we find
 \beq
 1.1 \times 10^{-4} \leq \Im \lambda_t \leq 1.9 \times 10^{-4}
 \eeq 
 in the first quadrant and
 \beq
 0.75 \times 10^{-4} \leq \Im \lambda_t \leq 1.9 \times 10^{-4}
 \eeq 
 in the second quadrant. For the range of $\hat B_K$ 
given in \eq{BKrange} varying all the other
parameters (including $m_t$ and $\Lambda_{\rm QCD}$) 
affects the above limits on $\Im \lambda_t$ by less than 20\%. 
In particular, the
upper bound on $\Im \lambda_t$ is stable and it is directly related 
to the maximum value of $\eta$ obtained from \eq{bound2} ($\rho=0$).
The upper bound
on $\Im \lambda_t$ becomes a sensitive function of the input parameters
only if we consider $\hat B_K > 0.5$.
In other words, we agree with ref.~\cite{PP} that
it is the theoretical uncertainty on the hadronic  
$\Delta S=2$ matrix element 
that controls the uncertainty on the determination
of $\Im \lambda_t$. 

We have included the bounds provided by \eqs{bound1}{bound2}
in all the following tables and figures.

\subsection{$\gamma_5$-scheme Independence}

 In order to fix $M$, we compare the computation in the HV $\gamma_5$-scheme with 
 that in the NDR. Figs. 1, 2 and 3 show how the 
 intersection between the two results remains stable
 as we change
 the value of the quark condensate. 
Fig. 4 shows the change in stability that occurs
 as we change $\Lambda_{\rm QCD}^{(4)}$.

\begin{figure}[ht]
\epsfxsize=10cm
\centerline{\epsfbox{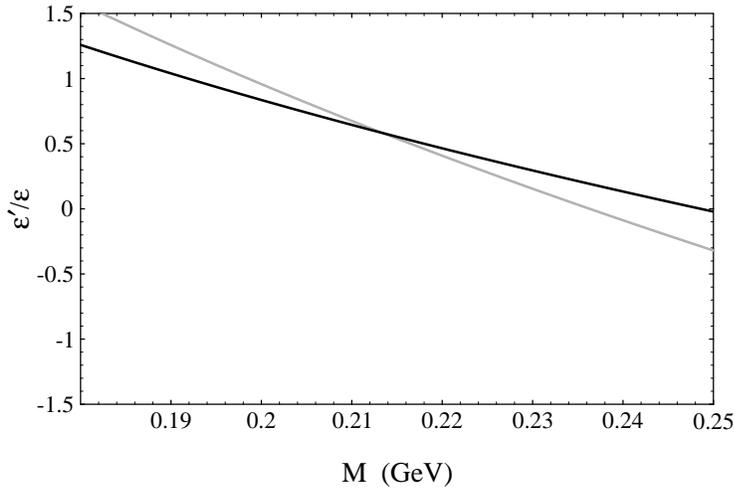}}
\caption{$\gamma_5$-scheme dependence of
$\varepsilon'/\varepsilon$. The black (gray) line represents 
the HV (NDR) result. The results are shown for
$\vev{\bar{q}q} (0.8\ {\rm GeV}) =( -200$ MeV)$^3$, 
$m_t^{\rm pole} = 180$ GeV,
$\Im \lambda_t = 1.3 \times 10^{-4}$
and $\Lambda_{\rm QCD}^{(4)} = 350$ MeV. The stability is obtained
at about $M = 215$ MeV. 
$\varepsilon'/\varepsilon$ is given in units of $10^{-3}$.}
\end{figure}

We find that the values at which $\gamma_5$-scheme independence is achieved
\beq
 M \simeq 215-220 \:\: \mbox{MeV}
 \eeq
 are quite stable with respect 
to different values of the matching, the quark and
 the gluon condensates and $m_t$. 
Smaller values of $M$ are selected
 for smaller values of
 $\Lambda_{\rm QCD}$ (and a correspondingly higher value of
 $\varepsilon'/\varepsilon$). These results 
are consistent with those  found in
 II for the $\Delta I = 1/2$ selection rule, 
where stability is achieved in the
 range $M = 160 - 220$. They are also consistent with the independent
 estimates discussed in section 3.

\begin{figure}[htb]
\epsfxsize=10cm
\centerline{\epsfbox{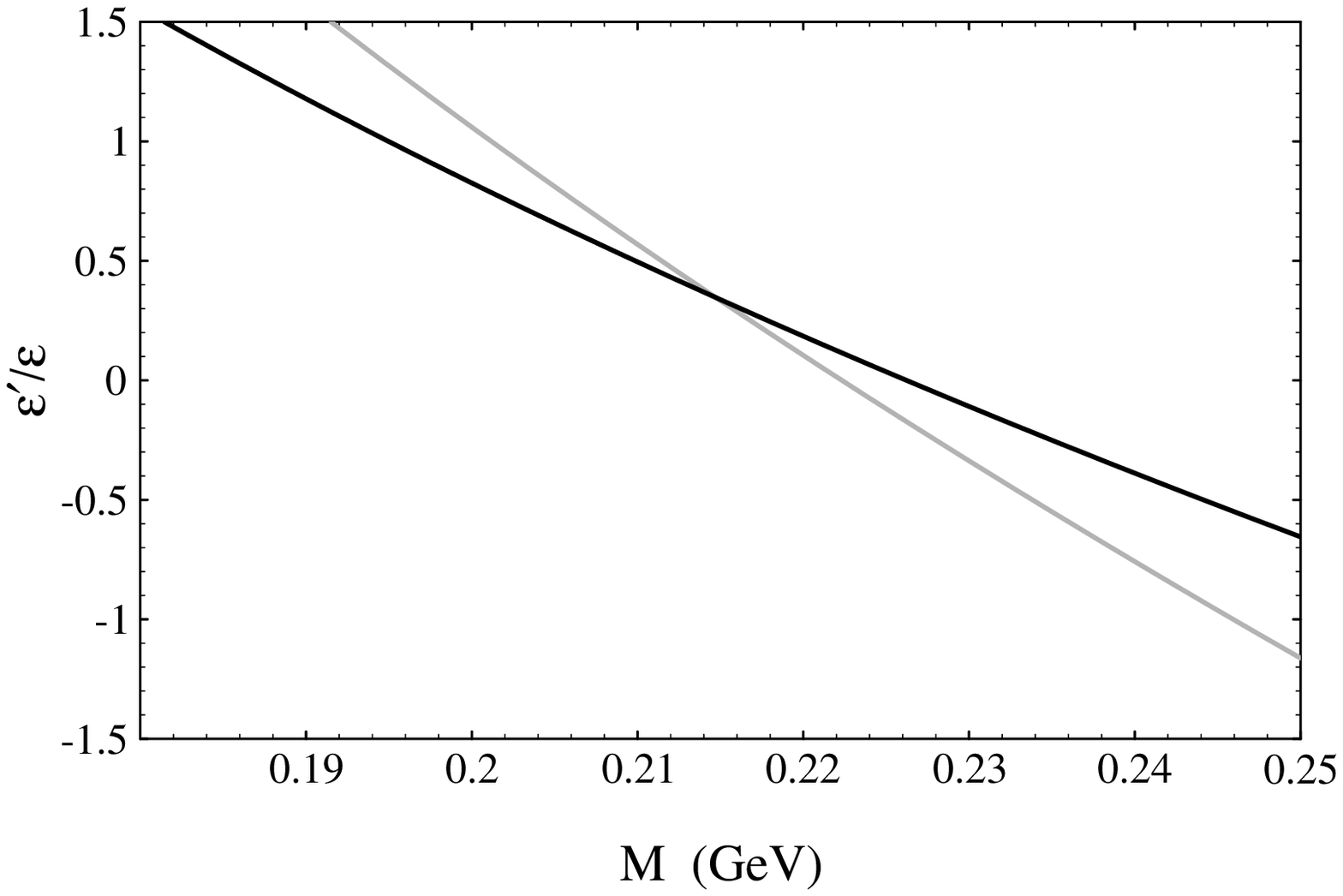}}
\caption{Same as Fig. 1 for 
$\vev{\bar{q}q} (0.8\ {\rm GeV}) =( -240$ MeV)$^3$.}
\end{figure}
\begin{figure}[htb]
\epsfxsize=10cm
\centerline{\epsfbox{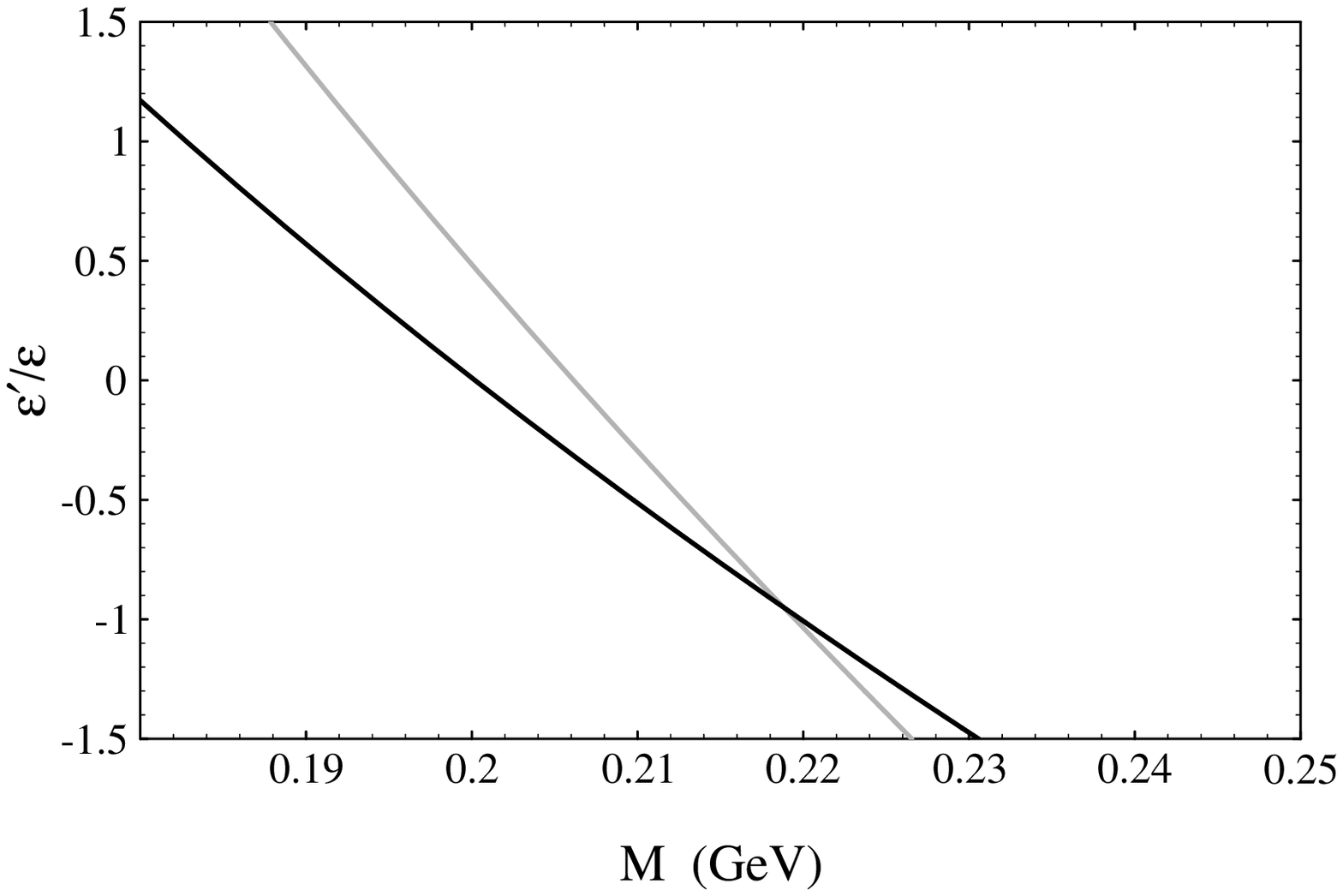}}
\caption{Same as Fig. 1 for 
$\vev{\bar{q}q} (0.8\ {\rm GeV}) =( -280$ MeV)$^3$.}
\end{figure}
\begin{figure}[htb]
\epsfxsize=10cm
\centerline{\epsfbox{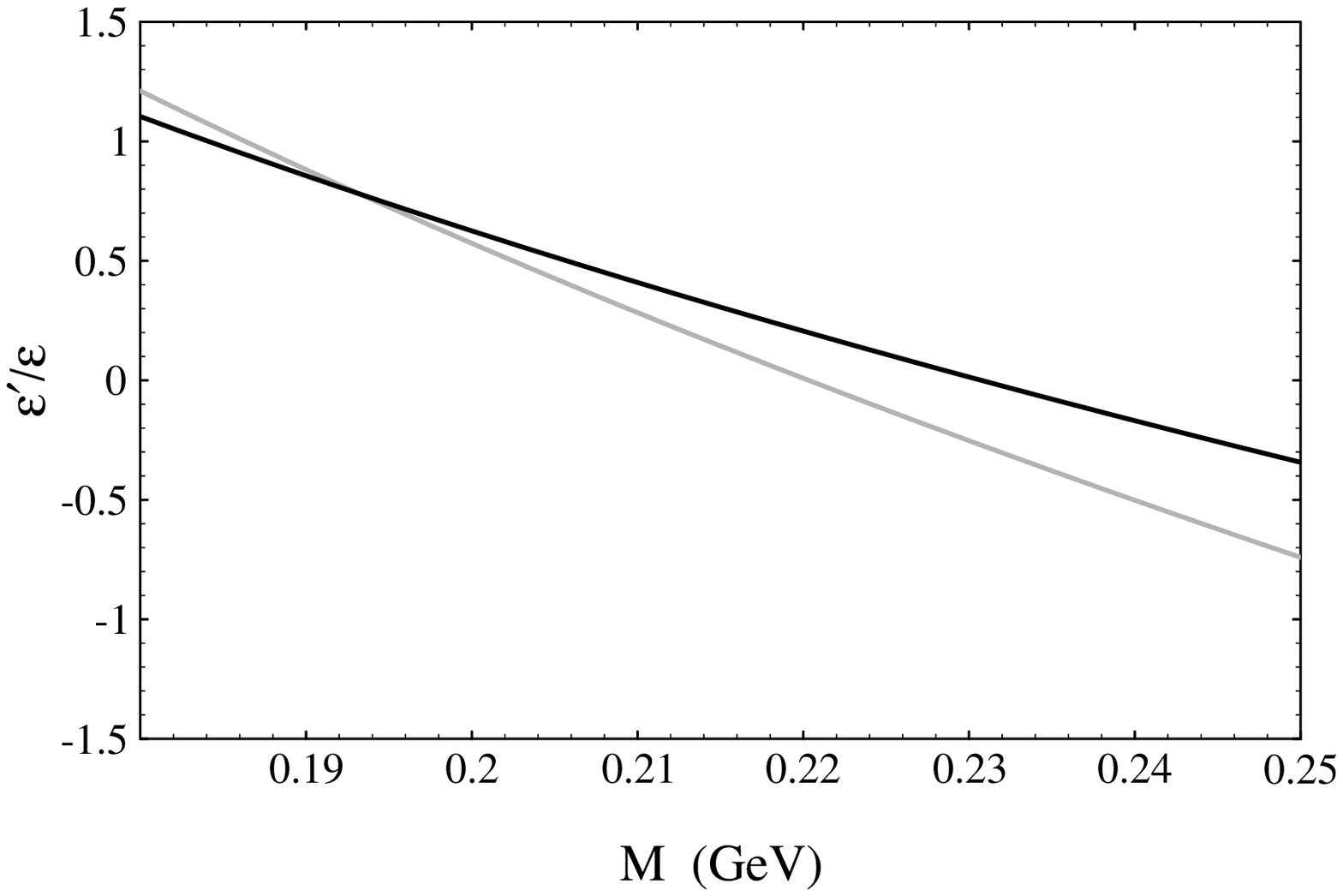}}
\caption{Same as Fig. 2 for $\Lambda_{\rm QCD}^{(4)} = 250$ MeV.}
\end{figure}
\begin{figure}[htb]
\epsfxsize=10cm
\centerline{\epsfbox{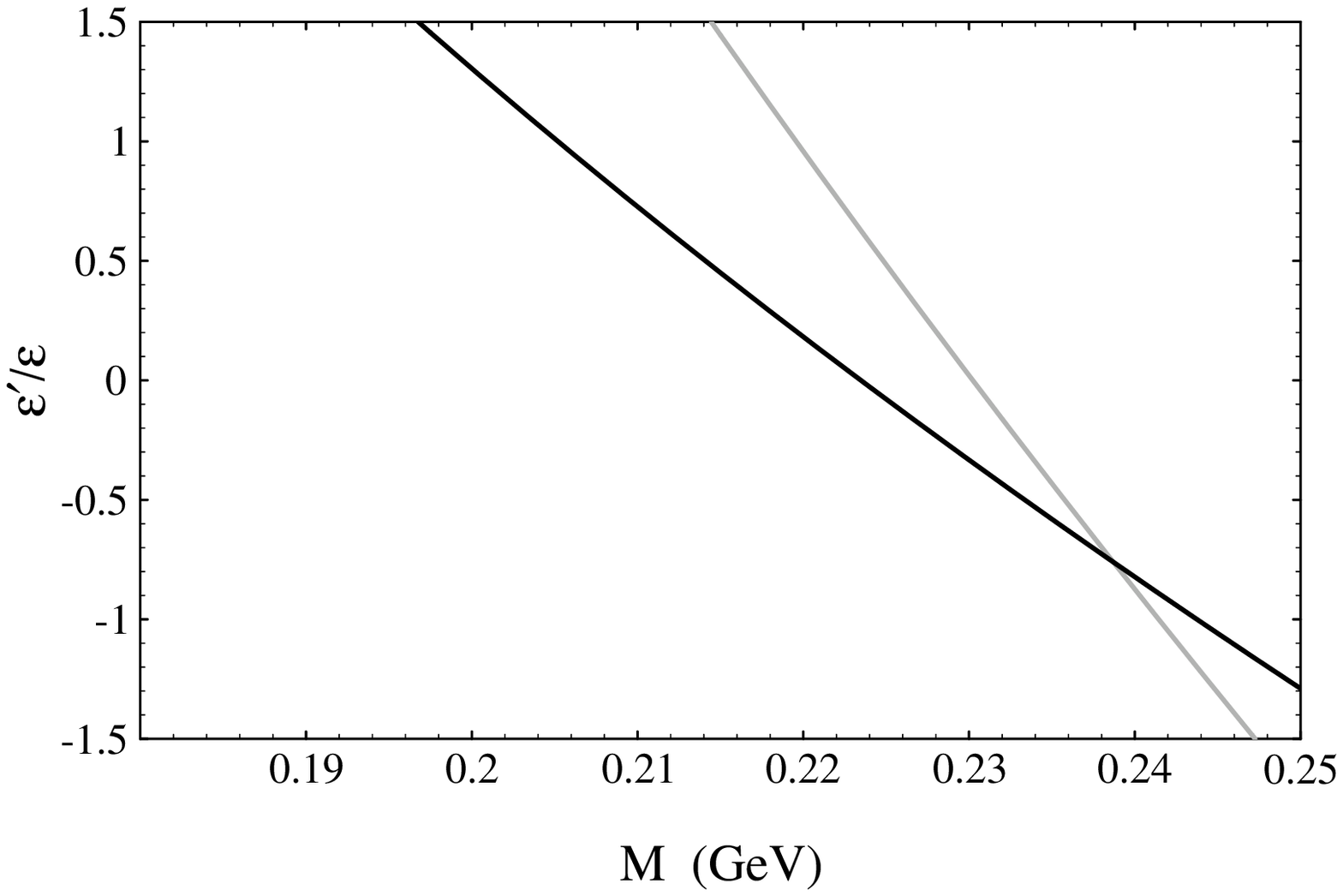}}
\caption{Same as Fig. 2 for $\Lambda_{\rm QCD}^{(4)} = 450$ MeV.}
\end{figure}

 As it is apparent from the figures, the final value of 
 $\varepsilon'/\varepsilon$ strongly depends on the value of $M$ we take. It is
 only through the device of requiring $\gamma_5$-scheme independence that we
 are able to reach a definite prediction. 
This procedure has the precious pay-off
of providing us with an improved estimate that does not suffer of the
uncertainty due to the $\gamma_5$-scheme dependence of the NLO Wilson
coefficients, which
may be as large as 80\%.
 
Figs. 4 and 5 show how the intersection depends on 
$\Lambda_{QCD}^{(4)}$.

\clearpage
\subsection{Anatomy of $\varepsilon'/\varepsilon$}

It is useful to consider the individual contribution
to $\varepsilon'/\varepsilon$
 of each
 of the quark operators. We have depicted them as histograms, where 
 the grey (black) one stands for the
 contribution before (after) meson-loop renormalization. Henceforth all
results are given for $M = 220$ MeV in the HV scheme.
 
It is clear from the 
 histograms of Fig. 6, 7 and 8 that the two 
 dominating operators are $Q_6$ and $Q_8$. Yet, since they give contributions
 approximately of the same size and opposite in sign, the final value turns out
 to be relatively small and of size comparable to that of most of
 the other operators. This result is at the origin the large theoretical
uncertainty as well as the unexpected smallness of  
$\varepsilon'/\varepsilon$.
\begin{figure}[ht]
\epsfxsize=10cm
\centerline{\epsfbox{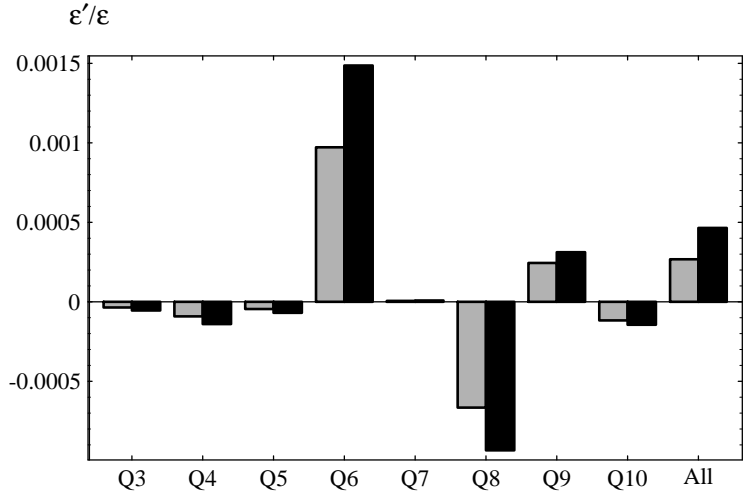}}
\caption{Histograms of the partial contributions to $\varepsilon'/\varepsilon$
of the height relevant operators for 
$\vev{\bar{q}q} (0.8\ {\rm GeV}) = (-200$ MeV)$^3$,
 $m_t^{\rm pole} = 180$ GeV,
$\Im \lambda_t = 1.3 \times 10^{-4}$ 
and $\Lambda_{\rm QCD}^{(4)} = 350$ MeV.
Gray (black) histograms represent the contribution of each operator 
without (with)
meson-loop renormalization. The last two histograms correspond to the sum of
all contributions.}
\end{figure}
\begin{figure}
\epsfxsize=10cm
\centerline{\epsfbox{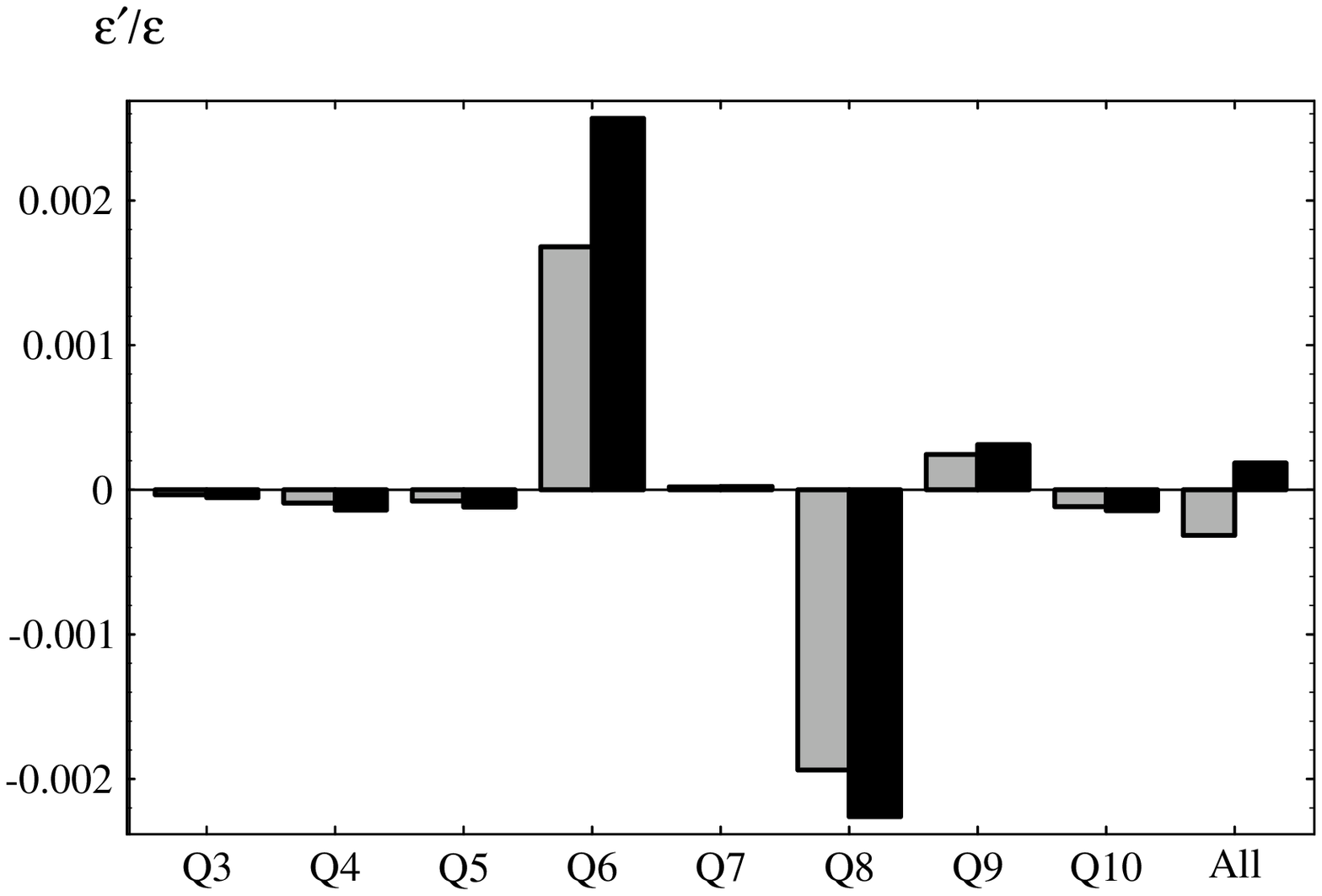}}
\caption{Same as Fig. 6 for 
$\vev{\bar{q}q} (0.8\ {\rm GeV}) = (-240$ MeV)$^3$.}
\end{figure}
\begin{figure}
\epsfxsize=10cm
\centerline{\epsfbox{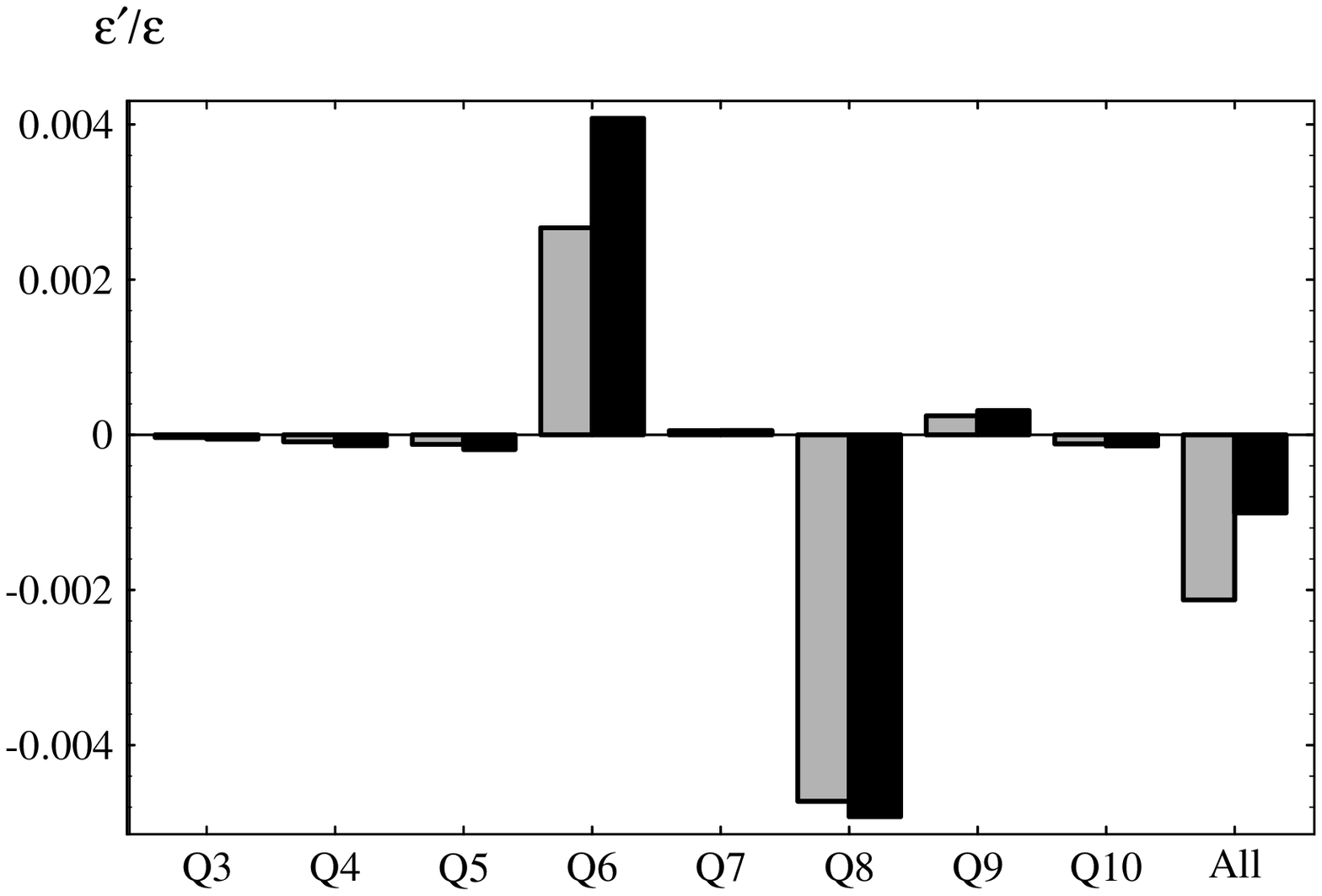}}
\caption{Same as Fig. 6 for 
$\vev{\bar{q}q} (0.8\ {\rm GeV}) = (-280$ MeV)$^3$.}
\end{figure}

The same histograms serve the purpose of showing that the meson-loop 
 renormalizations  are
 crucial not only in the overall size of each contribution but also
in determining the sign of the final result (see Fig. 7). 
These corrections are here consistently
included in the estimate for the first time.

The role of the operator $Q_4$ turns out to be
 marginal in our approach. In comparing 
this result with
that of the $1/N_c$ framework~\cite{buras} (see also
the final tables in ref. \cite{BEF} where we reproduce the individual $1/N_c$
contributions for the standard ten operators), it
should be recalled that in the above analysis
the $Q_4$ operator is written in terms of $Q_1$, $Q_2$ 
and $Q_3$ and that its values is therefore influenced by the $B_i$ factors
assigned to the former matrix elements. 
In particular, while $B_1$ and $B_2$ are  in ref.~\cite{buras} 
requested to be large
in order to account for the $\Delta I =1/2$ rule, $B_3$ 
is assigned the
value of 1. Such a procedure produces a rather large value for the
matrix element of $Q_4$. In our approach, we see that in fact also $B_3$ is 
large (and negative!) and that $Q_4$, 
once written in terms of the other operators,
is small, as found in the direct estimate.

\clearpage
\section{Estimating $\varepsilon'/\varepsilon$}

The preliminary work of the previous sections allows us to
estimate $\varepsilon'/\varepsilon$.  
The two most important sources of uncertainty are the quark condensate
  and the value
 of Im $\lambda_t$. Accordingly, we plot the values of 
 $\varepsilon'/\varepsilon$ as a function of
 these two quantities.
 Fig. 9 and 10 show our estimates, for $m_t$  
 fixed at its central value, in the first and second quadrant respectively.
 As it can be seen by inspecting these figures, the larger the value of the
 quark condensate, the swifter is the change in $\varepsilon'/\varepsilon$. 

\begin{figure}[ht]
\epsfxsize=10cm
\centerline{\epsfbox{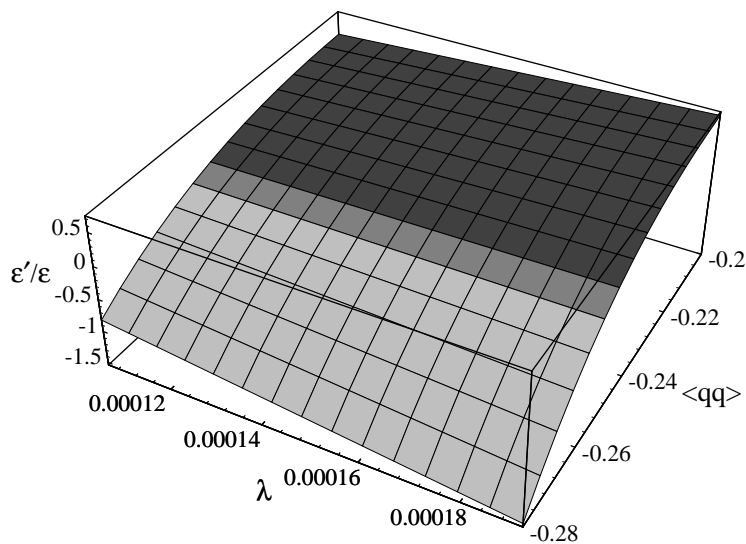}}
\caption{$\varepsilon'/\varepsilon$ in units of $10^{-3}$ for 
$m_t^{\rm pole} = 180$ GeV and $\Lambda_{QCD}^{(4)} = 350$ MeV
as a function of 
$\lambda \equiv \Im \lambda_t$ and
the quark condensate $\vev{qq} \equiv \vev{\bar q q}^{1/3}$ in units of GeV. 
Im $\lambda_t$ is taken in the first quadrant.
Black (grey) squares represent positive (negative) values.}
\end{figure}
\begin{figure}
\epsfxsize=10cm
\centerline{\epsfbox{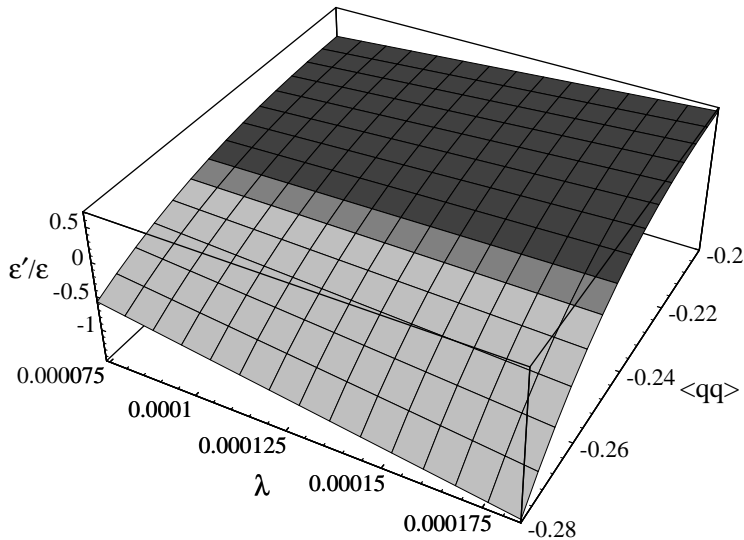}}
\caption{Same as Fig. 9 with Im $\lambda_t$ in quadrant II.}
\end{figure}
To have an  idea of the effect of varying $m_t$, the third major source of
uncertainty in the input parameters,  we have included Fig. 11 where
the top mass is varied in the given range. 

Fig. 12
shows the stability of our prediction for different matching scales 
$\mu = 0.8$ and 1 GeV (the perturbative running of $\vev{\bar q q}$ is 
included by taking the value of the condensate at $\mu = 0.8$ GeV as the
input value and than running it to $\mu =1$ GeV). The matching-scale
dependence is below 20\% in most of the range, becoming almost 30\%
only for very large values of the quark condensate.

\begin{figure}
\epsfxsize=10cm
\centerline{\epsfbox{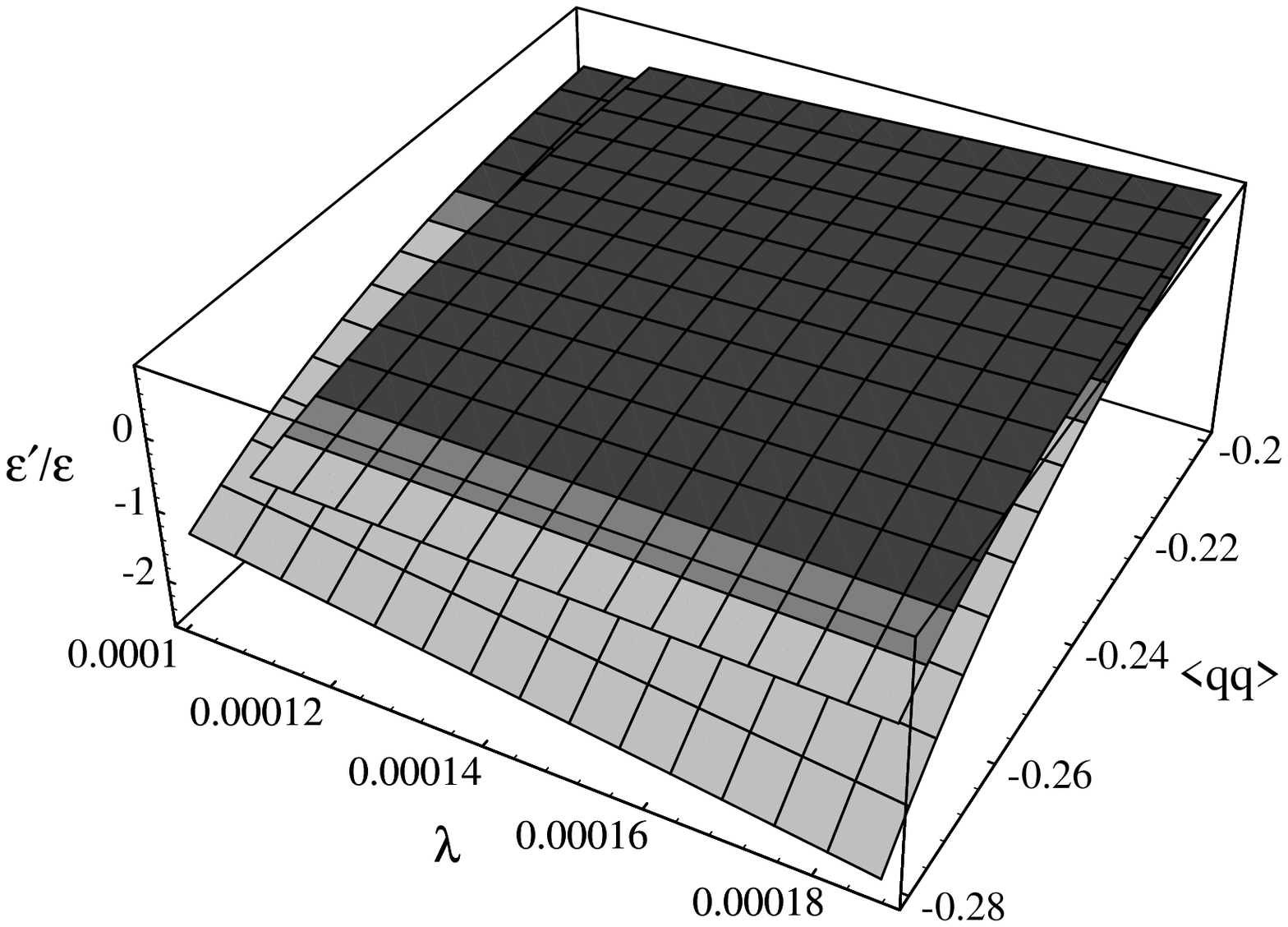}}
\caption{Same as Fig. 9 for two different values of $m_t$:
$m_t^{\rm pole} = 168$ GeV (upper surface) and 192 GeV (lower surface).}
\end{figure}
\begin{figure}
\epsfxsize=10cm
\centerline{\epsfbox{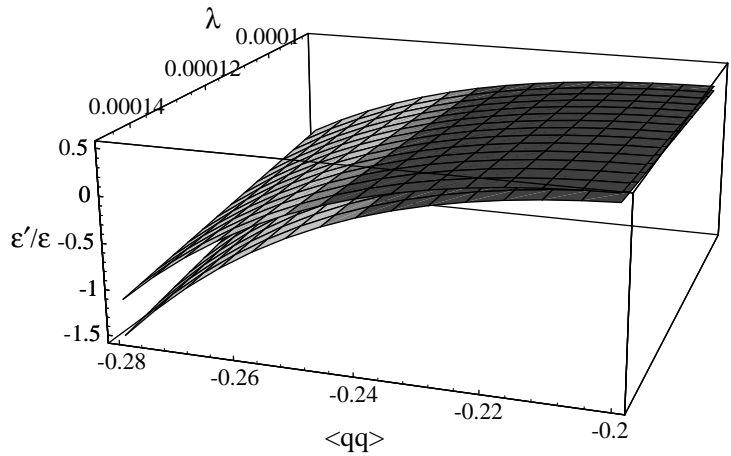}}
\caption{Same as Fig. 9 for different matching scales:
$\mu= 0.8$ GeV (upper surface) and 1.0 GeV (lower surface).}
\end{figure}
In order to provide the reader with a more analytical view, we have 
also collected in
Table 4 the numerical results at the varying of all the relevant parameters.

\begin{table}
\begin{center}
\begin{footnotesize}
\begin{tabular}{|c|c|c|c|}
\hline
\multicolumn{4}{|c|}{\mbox{$\Lambda_{\rm QCD}^{(4)} = 250$ Mev}}\\
\hline
\hline
$\vev{\bar{q}q}^{(1/3)}$ (MeV) & $m_t^{\rm pole}$ (GeV) & quadrant I & quadrant II \\
\hline
        & 168 & $3.7 \div 6.5$ & $2.7 \div 6.3$ \\
\cline{2-4}
 $-200$ & 180 & $2.9 \div 5.3$ & $2.0 \div 4.9$ \\
\cline{2-4}
        & 192 & $2.1 \div 3.9$ & $1.4 \div 4.1$ \\
 \hline
        & 168 & $3.5 \div 6.3$ & $2.6 \div 6.1$ \\
 \cline{2-4}
 $-240$ & 180 & $1.6 \div 3.0$ & $1.2 \div 2.8$ \\
 \cline{2-4}
        & 192 & $ -0.2 \div -0.4$ & $-0.1 \div -0.4$  \\
 \hline
        & 168 & $-0.1 \div -0.2$ & $-0.1 \div -0.2$ \\
 \cline{2-4}
 $-280$ & 180 & $-4.1 \div -7.5$ & $-2.9 \div -7.0$ \\
 \cline{2-4}
        & 192 & $-8.0 \div -15$ & $-0.5 \div -16$ \\
 \hline
 \hline
 \multicolumn{4}{|c|}{\mbox{$\Lambda_{\rm QCD}^{(4)} = 350$ Mev}}\\
 \hline
 \hline
 $\vev{\bar{q}q}^{(1/3)}$ (MeV) & $m_t^{\rm pole}$ (GeV) & quadrant I & quadrant II \\
 \hline
        & 168 & $5.0 \div 8.5$ & $3.7 \div 8.7$ \\
 \cline{2-4}
 $-200$ & 180 & $3.8 \div 6.9$ & $2.7 \div 6.6$ \\
 \cline{2-4}
        & 192 & $2.6 \div 4.7$ & $1.7 \div 4.9$ \\
 \hline
        & 168 & $4.3 \div 7.4$ & $3.2 \div 7.6$ \\
 \cline{2-4}
 $-240$ & 180 & $1.5 \div 2.7$ & $1.1 \div 2.6$ \\
 \cline{2-4}
        & 192 & $-1.3 \div -2.4$ & $-0.9 \div -2.5$ \\
 \hline
        & 168 & $-2.3 \div -4.0$ & $-1.7 \div -4.1$ \\
 \cline{2-4}
 $-280$ & 180 & $-8.2 \div -15$ & $-5.7 \div -14$ \\
 \cline{2-4}
        & 192 & $-14 \div -26$ & $-9.4 \div -27$ \\
 \hline
 \hline
 \multicolumn{4}{|c|}{\mbox{$\Lambda_{\rm QCD}^{(4)} = 450$ Mev}}\\
 \hline
 \hline
 $\vev{\bar{q}q}^{(1/3)}$ (MeV) & $m_t^{\rm pole}$ (GeV) & quadrant I & quadrant II \\
 \hline
        & 168 & $8.4 \div 14$ & $6.2 \div 14$\\
 \cline{2-4}
 $-200$ & 180 & $6.2 \div 11$ & $4.4 \div 11$ \\
 \cline{2-4}
        & 192 & $4.0 \div 7.4$ & $2.7 \div 6.9$ \\
 \hline
        & 168 & $6.5 \div 11$ & $4.8 \div 11$\\
 \cline{2-4}
 $-240$ & 180 & $1.5 \div 2.7$ & $1.1 \div 2.6$ \\
 \cline{2-4}
        & 192 & $-3.4 \div -6.2$ & $-2.3 \div -5.8$ \\
 \hline
        & 168 & $-6.9 \div -12$ & $-5.1 \div -11$ \\
 \cline{2-4}
 $-280$ & 180 & $-17 \div -30$ & $-12 \div -30$ \\
 \cline{2-4}
        & 192 & $-27 \div -50$ & $-18 \div -47$ \\
 \hline
\end{tabular}
\end{footnotesize}
\end{center}
\caption{Table of $\varepsilon'/\varepsilon$ in units of $10^{-4}$. 
Matching scale $\mu = 0.8$ GeV. The
two values corresponds to, respectively, the lower and upper bounds
of $\Im \lambda_t$, which are determined by consistency with $\varepsilon$.}
\end{table}
After twelve figures and four tables, we hope to have convinced the reader that
the quantity $\varepsilon'/\varepsilon$ is difficult to estimate with 
great precision. 
We think that
only the order of magnitude can be predicted in a completely reliable manner.
The reason is very simple: the final value is the result of the cancellation
between two, approximately equal in size, contributions. Accordingly, even
a small
uncertainty will be amplified and we are 
unfortunately  dealing with
rather large ones. And yet, the shear importance of this quantity impels us
to provide the best estimate we can.

By varying all parameters in the allowed ranges and, in
particular, 
taking the quark condensate---which is the major source of
uncertainty---between $(-200\ {\rm MeV})^3$ and $(-280\ 
{\rm MeV})^3$
we find
\beq
-27\times 10^{-4}\ < \varepsilon '/\varepsilon < \ 9 \times 10^{-4}\, ,
\eeq
where we have kept $\Lambda_{\rm QCD} ^{(4)}$ fixed at its central value. A larger
range, 
\beq
-50\times 10^{-4}\ < \varepsilon '/\varepsilon < \ 14 \times 10^{-4}\, ,
\label{grande}
\eeq
is obtained by varying $\Lambda_{\rm QCD}^{(4)}$  as well.

It should be stressed that the large range of negative values that we obtain
is a consequence of two characteristic features of our matrix elements:
i) the enhancement of the size of the electroweak matrix elements 
$\vev{Q_{8,7}}$ 
due to the coherent effects of the 
additional $O(p^2)$ contributions so far neglected 
(see discussion in sect. 5) and the chiral loop corrections; 
ii) the linear dependence 
on $\vev{\bar q q}$ of the leading gluon penguin matrix elements
compared to the quadratic dependence of the leading terms in the electroweak
matrix elements, which makes the latter prevail for large values
of the quark condensate. 
The effect of i) represents 
an enhancement of the leading electroweak matrix elements
by a factor two with respect to
the vacuum insertion approximation and present $1/N_c$ estimates 
(see table 3), while feature ii) is absent in the $1/N_c$ approach,
the quark condensate dependence being always quadratic.  

To provide a somewhat more restrictive estimate 
we may assume for the quark condensate the improved PCAC result, namely 
$\vev{\bar qq} = -(221\: \pm 17\ {\rm MeV})^3$ at our matching
scale $\mu = 0.8$ GeV,
 and thus find
\beq
\varepsilon '/\varepsilon  = \left\{ \begin{array}{ll}
 4.5 \: ^{+4.1}_{-5.4} \,\times \,10^{-4} & {\rm quadrant \: I} \\
 3.9 \: ^{+5.0}_{-4.5} \,\times \,10^{-4} & {\rm quadrant \: II} \, .
 \end{array} \right. 
 \label{best} 
\eeq
The value of  
$\varepsilon '/\varepsilon = ( 4 \pm 5) \times 10^{-4} $ quoted in the abstract
is obtained by averaging over the two quadrants in \eq{best}.

The range (\ref{rangeIV}) for the quark condensate, on which the above
 estimate is based, is not the favorite  one 
 by our analysis of the
 $\Delta I = 1/2$ selection 
rule in the $\chi$QM. The upper half of the more conservative
range (\ref{qqexp}) seems
to accommodate more naturally the rule, 
at least for a constituent mass $M \simeq
 220$ MeV---the value we find by requiring $\gamma_5$-scheme independence of
 $\varepsilon'/\varepsilon$.
For large values of the quark condensate the central values of
$\varepsilon'/\varepsilon$ shift toward the superweak regime,
and the role of meson loop corrections becomes crucial.

By taking the quark condensate in the 
range (\ref{rangeII}), the QCD-SR improved PCAC result,
we find
\beq
\varepsilon '/\varepsilon  = \left\{ \begin{array}{ll}
 1.4 \: ^{+6.5}_{-5.5} \,\times \,10^{-4} & {\rm quadrant \: I} \\
 1.2 \: ^{+9.3}_{-4.0} \,\times \,10^{-4} & {\rm quadrant \: II} \, .
 \end{array} \right.
 \label{secondbest} 
\eeq
Actually, for such a range of $\vev{\bar q q}$, negative 
central values of $\varepsilon '/\varepsilon$ in both quadrants
are obtained due to the extra terms of
the bosonization of the electroweak operators $Q_7$ and $Q_8$ 
neglected in the previous estimates. Only after the inclusion of the
meson-loop renormalization $\varepsilon '/\varepsilon$ turns  
to the positive central values of \eq{secondbest}.
\begin{figure}[htb]
\epsfxsize=10cm
\centerline{\epsfbox{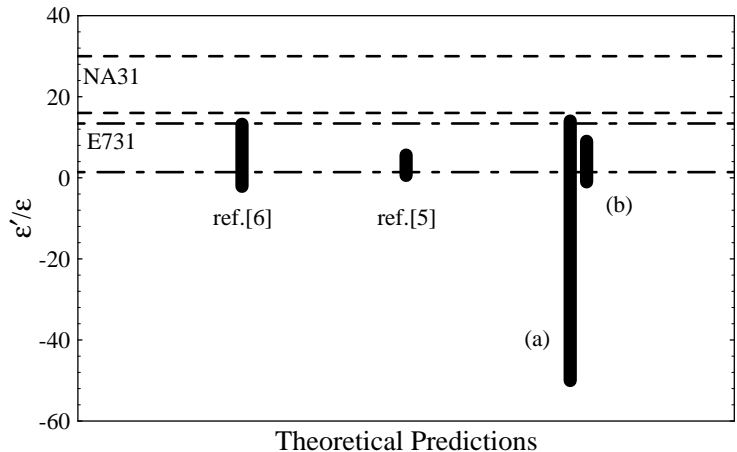}}
\caption{Present status of theoretical predictions and experimental values
for $\varepsilon'/\varepsilon$ (in units of $10^{-4}$). 
The most recent $1/N_c$ [6] and lattice [5] estimates are compared
to (a) our unbiased estimate (7.2), 
(b) our more restrictive estimate (7.3).}
\end{figure}
In Fig.13 we have summarized the present status of the theoretical predictions
for $\varepsilon'/\varepsilon$, compared to the present 1 $\sigma$ 
experimental results. 

\section{Outlook}

Our phenomenological analysis, 
based on the simplest implementation
of the $\chi$QM  and chiral lagrangian methods,
takes advantage of the observation that the $\Delta I = 1/2$ selection
rule in kaon decays is well reproduced in terms of three basic 
parameters (the constituent quark mass $M$ and the quark and gluon 
condensates) in terms of which all hadronic matrix elements
of the $\Delta S=1$ lagrangian can be expressed.

We have used the best fit of the selection rule to constrain the allowed
ranges of $M$, $\vev{\bar q q}$ and $\vev{GG}$ and we have fed them
in the analysis of $\varepsilon '/\varepsilon$. 

Nonetheless,
the error bars on the prediction of $\varepsilon'/\varepsilon$
remain large.
This is due to two conspiring features: 1) the destructive interference
between the large hadronic matrix elements of $Q_6$ and $Q_8$ which 
enhances up to an order of magnitude any related uncertainty in the final 
prediction (this feature is general and does not depend on the
specific approach); 2) the fact that large quark-condensate values
are preferred in fitting the isospin zero $K^0\to \pi\pi$ amplitude at $O(p^2)$
(which is a model dependent result).

Whereas little can be done concerning point 1) which makes 
difficult any theoretical attempt to predict $\varepsilon'/\varepsilon$ 
with a precision better than a factor two, an improvement on 2) can be 
pursued within the present approach. 

Two lines of research are in progress.
On the one hand, we are extending the analysis to $O(p^4)$ in the 
chiral expansion to gain  better
precision on the hadronic matrix elements and to determine in a 
self-consistent way the polinomial contributions from the chiral loops;
preliminary results indicate that the $\Delta I = 1/2$ rule is 
reproduced for smaller values of the gluon and quark condensates, thus reducing
our error bar, in the direction shown by our more restrictive estimate.
On the other hand, we are studying the $\Delta S = 2$ sector
to determine at the same order of accuracy
$\hat{B}_K$ and the $K_L$--$K_S$ mass difference by
including in the latter the interference with long-distance contributions 
that can be self-consistently computed in the present approach.

Whether this program is successfull may better determine how much of the 
long range dynamics of QCD is embedded in the present approach
and increase our confidence on the predictions of unknown observables.

\appendix
\section{Input Parameters} 
\begin{table}[htb]
\begin{center}
\begin{footnotesize}
\begin{tabular}{|c|c|}
\hline
{\rm parameter} & {\rm value} \\
\hline
$V_{ud}$ & 0.9753 \\
$V_{us}$ & $0.2205 \pm 0.0018$ \\
$\sin ^2 \theta_W$ & 0.2247 \\
$m_Z$ & 91.187 GeV \\
$m_W$ & 80.22 GeV \\
$m_b$ & 4.8 GeV \\
$m_c$ & 1.4 GeV \\
$|\varepsilon |$ & $(2.266 \pm 0.023) \times 10^{-3}$ \\
\hline
$ |V_{cb}|$ & $0.041 \pm 0.003$ \\
$|V_{ub}/V_{cb}|$ & $0.08 \pm 0.02$ \\
$m_t^{\rm pole}$ & $180 \pm 12$ GeV \\
$\widehat B_K$ & $0.55 \pm 0.25$ \\ 
\hline
$f_\pi = f_{\pi^+}$  &  92.4  MeV \\
$f_K = f_{K^+}$ & 113 MeV \\
$m_\pi = (m_{\pi^+} + m_{\pi^0})/2 $ & 138 MeV \\
$m_K = m_{K^0}$ &  498 MeV \\
$m_\eta$ & 548 MeV \\
$\Lambda_\chi$ & $2 \sqrt{2}\ \pi f_\pi$ \\
$\Omega_{\eta+\eta'}$ & $0.25\pm 0.05$ \\
\hline
$\Lambda_{QCD}^{(4)}$ & $350 \pm 100$ MeV \\
$\overline{m}_u + \overline{m}_d$ (1 GeV) & $12 \pm 2.5$ MeV \\
$\overline{m}_s$ (1 GeV) & $178 \pm 18$ MeV \\
$\vev{\bar{q}q}$  &  $- (200 \div 280 \: \mbox{MeV} )^3$ \\
$ \langle \alpha_s GG/\pi \rangle $ & $(376 \pm  47 \: 
\mbox{MeV} )^4 $ \\
\hline
\end{tabular}
\end{footnotesize}
\end{center}
\caption{Table of the numerical values of the input parameters.}
\end{table}

%
\clearpage
\renewcommand{\baselinestretch}{1}

\end{document}